\documentclass[%
reprint,
superscriptaddress,
amsmath,amssymb,
 aps, pra,
longbibliography
]{revtex4-2}
\usepackage[utf8]{inputenc}
\usepackage{newtxtext}   
\usepackage{newtxmath}   
\usepackage{natbib}
\usepackage{graphicx}
\usepackage{dcolumn}
\usepackage{bm}
\usepackage{hyperref}
\hypersetup{%
    bookmarksopen=false,
    bookmarksnumbered=true,
    pdfnewwindow=true,
    unicode=false,
    colorlinks=true,%
    citecolor=blue,
    linkcolor=black,
    urlcolor=blue,
    filecolor=blue
    }

\usepackage[caption=false]{subfig}
\usepackage{mathrsfs}
\usepackage{array}
\usepackage{mathtools}
\usepackage{braket}
\usepackage[dvipsnames]{xcolor}
\definecolor{Blue}{rgb}{0.3,0.3,0.9}
\definecolor{Red}{rgb}{0.9,0.3,0.3}
\definecolor{Green}{rgb}{0.3,0.6,0.3}

\newcommand{\revisionRAR}[1]{{{#1}}}

\usepackage{neuralnetwork}
\usepackage{ifthen}
\usepackage{textcomp}

\newif\ifNOSUP \NOSUPtrue

\newcommand{\HubNet}{\textsc{HubbardNet}}

\begin{document}

\title{
Putting machine learning to the test in a quantum many-body system
}

\author{Yilun Gao}
\affiliation{%
Department of Physics, University of Warwick, Gibbet Hill Road, Coventry, CV4 7AL, UK
}%
 
\author{Alberto Rodr\'iguez}
\affiliation{%
Departamento de F\'isica Fundamental, Universidad de Salamanca, E-37008 Salamanca, Spain
}%
\affiliation{%
Instituto Universitario de F\'isica Fundamental y Matem\'aticas (IUFFyM), Universidad de Salamanca, E-37008 Salamanca, Spain
}%
\author{Rudolf A. R\"{o}mer}%
\affiliation{%
Department of Physics, University of Warwick, Gibbet Hill Road, Coventry, CV4 7AL, UK
}%

\date{\today}
             
\begin{abstract}
Quantum many-body systems pose a formidable computational challenge due to the exponential growth of their Hilbert space. While machine learning (ML) has shown promise as an alternative paradigm, most applications remain at the proof-of-concept stage, focusing narrowly on energy estimation at the lower end of the spectrum. Here, we push ML beyond this frontier by \revisionRAR{extensively} testing \HubNet, a deep neural network architecture for the Bose-Hubbard model. Pushing improvements in the optimizer and learning rates, and introducing physics-informed output activations that can resolve extremely small wave-function amplitudes, we achieve ground-state energy errors reduced by orders of magnitude and wave-function fidelities exceeding 99\%. We further assess physical relevance 
by analysing generalized inverse participation ratios and multifractal dimensions for ground and excited states in one and two dimensions, demonstrating that optimized ML models reproduce localization, delocalization, and multifractality trends across the spectrum. 
Crucially, these qualitative predictions remain robust across four decades of the interaction strength, e.g.\ spanning across superfluid, Mott-insulating, as well as quantum chaotic regimes.  
Together, these results suggest ML as a viable \emph{qualitative} predictor of many-body structure, complementing the quantitative strengths of exact diagonalization and tensor-network methods.
\end{abstract}

\keywords{}
\maketitle

\section{\label{sec:Introduction} Introduction}

Machine learning (ML) has recently emerged as a flexible, optimization-driven paradigm for tackling some of the most challenging problems in theoretical and computational physics \cite{Alpaydin2020,Hinton1999,Carleo2019,Mehta2019}, particularly those involving quantum many-body systems \cite{Schollwck2011,Tasaki2020,Fazio2025}. These systems exhibit an exponentially growing Hilbert space with system size, making traditional approaches such as exact diagonalization (ED) feasible only for small systems \cite{Feynman1982,Troyer2005,Jung2020,Gao2025SpectralChain}. While tensor network methods like DMRG \cite{Ostlund1995,Schollwck2011,Verstraete2023} and PEPS \cite{Orus2014,Schuch2007} have extended our reach, they too face limitations in higher dimensions or in systems with strong entanglement and disorder \cite{Kottmann2022}. In particular, these methods are not designed for rapid parameter sweeps.

Early applications of ML in physics have largely demonstrated proof-of-principle successes, showing that neural-network-based representations of quantum states can capture essential features of ground states \cite{Carleo2017,Nomura2017} and, in some cases, outperform conventional variational methods \cite{Vivas2022,Kottmann2022}. 
Still, systematic tests of wave-function fidelity across distinct interaction regimes are scarce.
Beyond proof-of-concept energy and ground state estimates \cite{McClean2016,Saito2017,Saito2018,Hibat-Allah2020}, a key question remains:  Can ML reproduce the structure of many-body eigenstates 
across phase transitions and deep in the excitation spectrum? This question is central to the connection between eigenstate structure and experimentally relevant observables in regimes governed by quantum chaos 
and the Eigenstate Thermalization Hypothesis (ETH) \cite{Borgonovi2016, DAlessio2016, Deutsch2018, Trotzky2012, Santos2012, Rigol2008}.

Here we stress-test the recently proposed \HubNet\ \cite{Zhu2023} on the Bose-Hubbard model (1D and 2D) \cite{Fisher1989,Lewenstein2007,Bloch2008, Bloch2012, Krutitsky2016,Cazalilla2011}. With optimizer and learning-rate improvements and physics-informed output activations, we achieve sub-percent ground-state energy errors, $>99\%$ wave-function fidelity, and robust predictions of generalized fractal dimensions (GFD) \cite{Mirlin2000,Evers2008,Rodriguez2011} over four decades in the interaction strength parameter. 
This achievement goes far beyond previous results, showing that a single configuration of a neural network can interpolate between two very distinct physical regimes separated by a quantum phase transition.
Furthermore, for excited states, we introduce \emph{observable-based training} employing \revisionRAR{GFD}  
that bypasses \revisionRAR{the need for} Gram-Schmidt \revisionRAR{orthogonalization} towers and yields accurate 
eigenstate-structure statistics in the spectrum bulk across quantum chaos regimes. We thus provide a comprehensive framework to establish how ML-based methods can capture the structural features of many-body states that underpin experimentally relevant observables. 

While we build on \HubNet\ as introduced in Ref.\ \cite{Zhu2023}, all results concerning multi-order-of-magnitude parameter sweeps,  observable-based training of excited states, and activation-function engineering are new to the present work.
Our results demonstrate that ML, when carefully optimized, can go well beyond proof-of-concept and offer a viable route to studying complex quantum systems. At the same time, we highlight the challenges that remain---particularly in scaling to larger systems and accurately describing states deep in the spectrum---thus outlining directions for future research.

Section \ref{sec:Model} defines the Bose-Hubbard model and structure metrics (the GFD). Section \ref{sec:Neural-Network} details \HubNet\ and our training improvements. Sections \ref{sec:ground-state}-\ref{sec:excited-states} present ground- and excited-state results, and Sec.~\ref{sec:discussions-and conclusions} discusses implications and limitations.

\section{\label{sec:Model} The Bose-Hubbard Model}

\subsection{\label{subsec:model-definitions} Definition and notation}

We study $N$ interacting bosons on a $d$-dimensional lattice described by the Bose-Hubbard Hamiltonian (BHH) \cite{Fisher1989}, 
\begin{equation}
    H=-J\sum_{\langle i,j \rangle}\hat{a}_i\hat{a}^{\dagger}_j+\frac{U}{2}\sum_{i=1}^{M}\hat{n}_i(\hat{n}_i-1),
    \label{eq:BHH}
\end{equation}
where $\hat{a}_i (\hat{a}_i^{\dagger})$ are annihilation (creation) operators in single particle Wannier states, localized at the different lattice sites $i$, $\hat{n}_i$ being the corresponding number operator, $J$ is the energy associated with particle tunneling between nearest neighboring sites ${\langle i,j \rangle}$, $U$ represents the onsite pair interaction energy, and $M$ stands for the total number of sites. In the following, we always take the tunneling energy as the energy unit, i.e., $J=1$, and periodic boundary conditions are assumed.

The BHH models a paradigmatic quantum many-body system \cite{Krutitsky2016,Cazalilla2011} that can be faithfully implemented experimentally using ultracold atoms in optical potentials \cite{Lewenstein2007,Bloch2008,Bloch2012}, and exhibits a rich phenomenology, including a ground state quantum phase transition between a Mott insulator and a superfluid, controlled by the ratio $J/U$ and the bosonic density, which has been experimentally observed \cite{Greiner2002,Boeris2016,Bakr2010}. Additionally, the complex excitation spectrum of the system gives rise to quantum chaos \cite{Pausch2021,Pausch2021b,Pausch2022}, inducing the emergence of dynamical ergodic behaviour \cite{Duenas2025,Pausch2025}. Hence, this bosonic model (and variations thereof) provides an ideal platform for the investigation of strongly correlated dynamics and thermalization in isolated quantum many-particle systems \cite{Srednicki1994, Srednicki1999,Borgonovi2016,DAlessio2016,Deutsch2018}, questions which are currently being intensively studied both theoretically and experimentally \cite{Cheneau2012,Trotzky2012,Ronzheimer2013,Choi2016,Lukin2019,Rispoli2019,Takasu2020,Leonard2023}.

In the one dimensional case (1D, $d=1$), one has a simple chain with $M$ sites, while in 2D ($d=2$), we assume a square lattice with $M = \sqrt{M} \times \sqrt{M}$ and $\sqrt{M}$ a positive integer. 
Since Hamiltonian \eqref{eq:BHH} conserves the total particle number $N$, the system can be independently diagonalized in Hilbert subspaces of dimension 
\begin{equation}
  \text{dim}\mathcal{H}=\binom{M+N-1}{N},
\end{equation}
which at fixed particle density grows exponentially with $N$ or $M$ [e.g., at unit density $\text{dim}\mathcal{H}(N+1)\approx 4\ \text{dim}\mathcal{H}(N)$].
In the cases $J=0$ 
(no tunneling) or $U=0$ (non-interacting), the model is integrable,  and the eigenstates can be uniquely identified by $M$ quantum numbers, due to the existence of $M$ observables that commute with the Hamiltonian. For $J=0$, the eigenstates are given by the Fock states
\begin{equation}
    \ket{f}\equiv\ket{n_{1},n_{2}....,n_{M}},
    \label{eq:basis}
\end{equation}
determined by the sequence of occupation numbers $n_i\geqslant 0$ of the lattice sites such that $\sum_i n_{i}=N$, and where each sequence can be uniquely identified by an integer index $f\in[1,\text{dim}\mathcal{H}$]. The latter set of Fock states spans a convenient basis in which to diagonalize the system in the general case with $J\neq0$ and $U\neq0$.
The 
$k$-th eigenstate $\ket{\Psi_k}$, ordered by increasing energy $E_k$, would then be written as 
\begin{equation}
    \ket{\Psi_k}=\sum_{f=1}^{\text{dim}\mathcal{H}}\psi_k(f)\ket{f}, \quad k=0,1 \ldots, \text{dim}\mathcal{H}-1,
    \label{eq:state}
\end{equation}
with corresponding expansion coefficients denoted by $\psi_k(f)$.
Our ML outputs will directly target these coefficients.

\subsection{Characterization of eigenstate structure}
\label{sec:model-states}

The two most relevant features of the BHH, namely its ground state phase transition and the emergence of a chaotic phase in its excitation spectrum, can be characterized by the accompanying pronounced changes in the structure of the many-particle eigenstates. The passage from the Mott insulating state to the superfluid as a function of $U$ correlates with a delocalization tendency of the ground state in Fock space that encodes the critical point \cite{Lindinger2019}, which in 1D 
at unit density is $U_c\approx 3.3$ \cite{Kashurnikov1996,Ejima2011,Carrasquilla2013,Boeris2016}. On the other hand, the chaotic phase is populated by eigenstates that become fully extended (ergodic) in Fock space in the thermodynamic limit \cite{Pausch2021,Pausch2021b,Pausch2022,MartnClavero2025}.

To quantify wave-function structure in the Fock basis, we use the finite-size
\revisionRAR{GFD}, $D_q$ \cite{Mirlin2000,Rodriguez2011}, which, for $\ket{\Psi_k}$ as given above, are defined as
\begin{equation}
D_q = \frac{1}{1-q}\frac{\text{ln} \left[ \sum_f |\psi_k(f)|^{2q} \right]} {\text{ln} (\text{dim}\mathcal{H})}, \quad q\in \mathbb{R}.
\label{eq:Dq}
\end{equation}
The values of $D_q$ as $\text{dim}\mathcal{H}\to\infty$ reveal the structural features of the state in the chosen basis for the expansion. In our case, $D_q$ probes how wave-function weight is distributed across the Hilbert space: 
A fully extended state is characterized by $D_q\to 1$ for any $q$, whereas $D_{q>1}\to 0$ signals localization in Fock space, i.e., emphasising rare, large-amplitude components. On the other hand, the convergence of $D_q$ toward $q$-dependent values different from $0$ and $1$ corresponds to multifractality \cite{Paladin1987}, a property that commonly appears for many-body states in Fock space \cite{Lindinger2019,Mace2019,Pausch2021} and that plays a significant role in disordered systems \cite{Evers2008,Rodriguez2011,Pino2020}. The dimensions $D_q$ encode the behaviour of the distribution of wave-function intensities, where different values of $q$ correlate with particular ranges of $|\psi_k(f)|^2$, e.g., small wave-function intensities determine $D_{q<0}$, whereas large intensities contribute most prominently to $D_{q>0}$.

Among all possible $D_q$, we single out 
\begin{equation}
    D_1  \equiv \lim_{q\to1} D_q = 
-\frac{\sum_f |\psi_k(f)|^{2} \ln|\psi_k(f)|^{2} }{\ln (\text{dim}\mathcal{H})}, 
\end{equation}
that converges to the dimension of the support set of the state \cite{DeLuca2013}, $D_2$ that determines the scaling of the inverse participation ratio $\sum_f |\psi_k(f)|^4$ with Hilbert space dimension, and 
\begin{equation}
 D_{\infty}=-\frac{\ln \left[ \max_f|\psi_k(f)|^{2} \right]}{\ln (\text{dim}\mathcal{H})},
\end{equation}
which is entirely determined by the maximum wave-function intensity. 

\begin{figure}[tb]
\centering
\includegraphics[width=0.95\columnwidth]{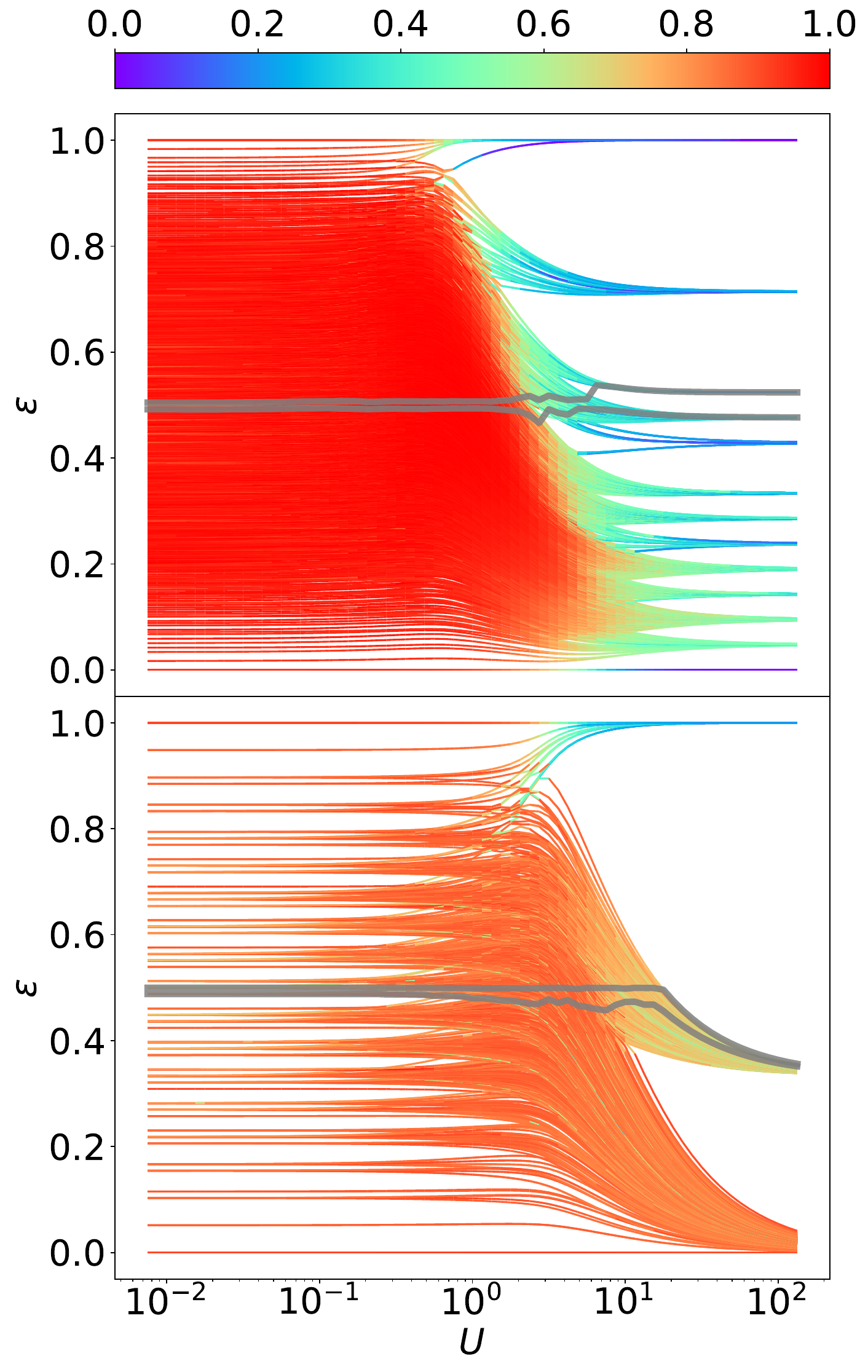} 
\caption{Variation of the fractal dimension $D_1$ (colour scale) as a function of interaction strength $U$ and scaled energy $\varepsilon$ for (top panel) all $1716$ states of a 1D BHH with system size $M=7$ and particle number $N=7$, and (bottom panel) a 2D BHH on a $4\times4$ square lattice with particle number $N=3$ for all $816$ states. The grey lines in both panels indicate the upper and lower bounds for the $50$ states closest to $\varepsilon=0.5$.
These trends serve as reference data for the ML predictions analysed in Sections \ref{sec:ground-state} and \ref{sec:excited-states}.
} 
\label{fig-1d-2d-Dq}
\end{figure}

An example of the information conveyed by the \revisionRAR{GDF} is given in Fig.~\ref{fig-1d-2d-Dq}, showing the evolution of $D_1$ as a function of $U$ for the entire spectrum of a BHH in 1D (upper panel) and 2D (lower panel) obtained by ED. 

The energy spectrum is represented in terms of the \emph{scaled energy} 
\begin{equation}
\varepsilon = \frac{E-E_0}{\max_k \left\{ E_k\right\} - E_0},
\end{equation}
for each value of $U$. 
The behaviour exhibited in the upper panel of Fig.\ \ref{fig-1d-2d-Dq} is exemplary for all $D_q$ \cite{Pausch2021,Pausch2021b} and unveils a change in the eigenstate structure associated with the emergence of a chaotic phase: For large interaction, the states are strongly localized in Fock space (low values of $D_1$), since they become the basis states \eqref{eq:basis}, and appear organized in degenerate manifolds corresponding to the different eigenenergies of the interaction term in the BHH, $E_k^{J=0}=(U/2)\sum_i n_i(n_i-1)$ (for $M=N=7$, these amount to 13 different values). As $U$ decreases, the degeneracy is lifted and at $U\approx 1$ there is a strong basis mixing yielding states that are markedly extended in Fock space ($D_q\to 1$). This region of delocalized states, corresponding to the chaotic phase, extends into the low interaction regime. 

While the existence of quantum chaos in the 2D BHH 
has not yet been properly investigated (due mainly to the very demanding numerical analysis), the behaviour of $D_1$ in the lower panel of Fig.~\ref{fig-1d-2d-Dq}, for a $4 \times 4 $ system with particle number $N=3$, also reveals an intermediate regime around $U\approx 2$ in the bulk of the spectrum populated by fairly delocalized states in Fock space. In this case, however, since particle density is much lower than unity ($N/M\simeq 0.19$), $69\%$ of the eigenstates converge to a non-interacting configuration in the limit $U\gg 1$ (i.e., with vanishing energy in that limit). Consequently, such highly degenerate manifold mixes a very large number of basis Fock states very quickly as $U$ decreases, giving rise to delocalized states much earlier as compared to the 1D case above.


\section{\label{sec:Neural-Network} The ML approach to the BHH}

At its core, ML is an optimization paradigm---an approach that aligns naturally with the variational principles of quantum mechanics \cite{Schrodinger1926}. 
Essentially, ML optimizes the coefficients of many-body states, i.e.\ when expressed in a Fock basis \cite{Carleo2017}. Several neural network architectures have been proposed to represent many-body quantum states, including \textsc{FermiNet} \cite{Spencer2021}, \textsc{PauliNet} \cite{Hermann2020}, \textsc{BosonNet} \cite{Denis2025}, and \HubNet\ \cite{Zhu2023}.
We choose \HubNet\ as a deliberately minimal, easily reproducible deep neural-network baseline---simple enough to implement (from the explicit code base on its github repository \cite{Zhu2023}) and audit, yet expressive enough to support a rigorous, end-to-end evaluation beyond energies.

\subsection{\label{sec:Neural-Network-structure} The principles of {\HubNet}}

\begin{figure}[tb]
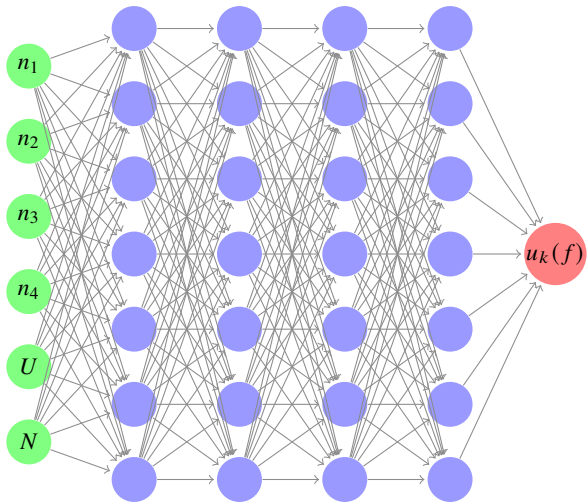

\begin{neuralnetwork}[height=5, layerspacing=14mm]
		\newcommand{\nodetextclear}[2]{}
		\newcommand{\nodetextx}[2]{%
  \ifthenelse{\equal{#2}{5}}{$U$}{%
    \ifthenelse{\equal{#2}{6}}{$N$}{$n_{#2}$}%
  }%
} 
        \newcommand{\nodetextz}[2]{$u_{k}(f)$}
		\inputlayer[count=6, bias=false, text=\nodetextx]
		\hiddenlayer[count=7, bias=false, text=\nodetextclear] \linklayers
        \hiddenlayer[count=7, bias=false, text=\nodetextclear] \linklayers
        \hiddenlayer[count=7, bias=false, text=\nodetextclear] \linklayers
        \hiddenlayer[count=7, bias=false, text=\nodetextclear] \linklayers
		\outputlayer[count=1, text=\nodetextz] \linklayers
\end{neuralnetwork}

\caption{The MLP structure of \HubNet\ with $M=4$. The green circles represent the input layer with the $n_i$'s, $U$, and $N$ as the input parameters, while the purple circles denote the $4$ hidden layers. The output neuron is indicated by the red circle.
}
\label{fig-nn}
\end{figure}

\HubNet\ is a fully-connected multilayer perceptron (MLP) \cite{DavidERumelhart1986} design as shown schematically in Fig.\ \ref{fig-nn}: Inputs are the occupation numbers $\{n_i\}_{i=1}^M$, the interaction strength $U$, and the particle number $N$. The neural network (NN) contains four hidden layers ($400$ neurons, $\tanh$) and produces a single real output $u_k(f)$  
\footnote{\citet{Zhu2023} use two neurons in the output layer corresponding to the real and imaginary parts of the wave-function. As the wave-functions for the BHH, with a real hopping constant, and without gauge fields, can always be chosen to be real, the use of a single output neuron is sufficient for our purposes.}. 
The latter output is converted into the wave-function coefficient associated with Fock state $\ket{f}$ [see Eqs.~\eqref{eq:basis} and \eqref{eq:state}] via an activation function $\sigma(u)$ plus normalization, 
\begin{equation}
\psi^{\textrm{NN}}_k(f)=\frac{\sigma[u_k(f)]}{\|\sigma[u_k]\|},\quad f=1, \ldots,  \text{dim}\mathcal{H},
\label{eq:output}
\end{equation}
where the superscript NN denotes neural-network prediction. 
By default, \HubNet\ minimizes the following sum of Rayleigh quotients, 
\begin{equation}
\mathcal{L}_k=\sum_\gamma\left( 
\frac{\bra{\Psi_k^{\text{NN}}}H_\gamma\ket{\Psi_k^{\text{NN}}}}{\braket{\Psi_k^{\text{NN}}|\Psi_k^{\text{NN}}}} - \mathcal{L}_P\right),
\label{eq:loss}
\end{equation}
where $\gamma$ stands for a collection of training points (e.g., multiple values of $U$ and/or of $N$), and 
$\mathcal{L}_P$ represents an orthogonality penalty used only for excited states ($k>0$) to discourage overlap with previously learned lower states \cite{Zhu2023}.
The MLP employs an ingenious cosine annealing scheme \cite{Loshchilov2017} with a resetting period of $1000$ steps in order to avoid getting stuck in local minima. 
For the ground state $\ket{\Psi_0}$, \HubNet\ uses an exponential activation function $\sigma\left(u\right)=\exp\left( u \right)$
in the output layer since the expansion coefficients $\psi_0(f)$ are always positive in the BHH.
For excited states, \citet{Zhu2023} choose a linear activation function, $\sigma(u)=u$, and find the desired target excited state $\ket{\Psi_k}$ by iteratively constructing the full tower of lower-lying states $\ket{\Psi_{k'}}$ with $k'= 0, 1,\ldots k-1$ and employing a Gram-Schmidt process to ensure orthogonality of the eigenspace. 
The computational bottleneck is the per-iteration Rayleigh quotient, which scales with the number of sampled Fock configurations. For $M=N=7$ ($\text{dim} \mathcal{H}=1716$), energies via ED are seconds-scale; ML requires $\gtrsim 10^4$ iterations but amortizes across $U$ once trained.
With this setup, \citet{Zhu2023} showed promising energy and state predictions for $U\in[1,15]$. 
We demonstrate robust generalization across four decades in $U$, high-fidelity wave-functions, and accurate eigenstate-structure metrics---and we introduce observable-based training for excited states that avoids Gram-Schmidt towers.

\subsection{\label{sec:ml} Putting \HubNet\ to the test}
\subsubsection{\label{sec:ml-training} Training \& out-of-data predictions}

\begin{figure}[t]
\includegraphics[width=0.48\textwidth]{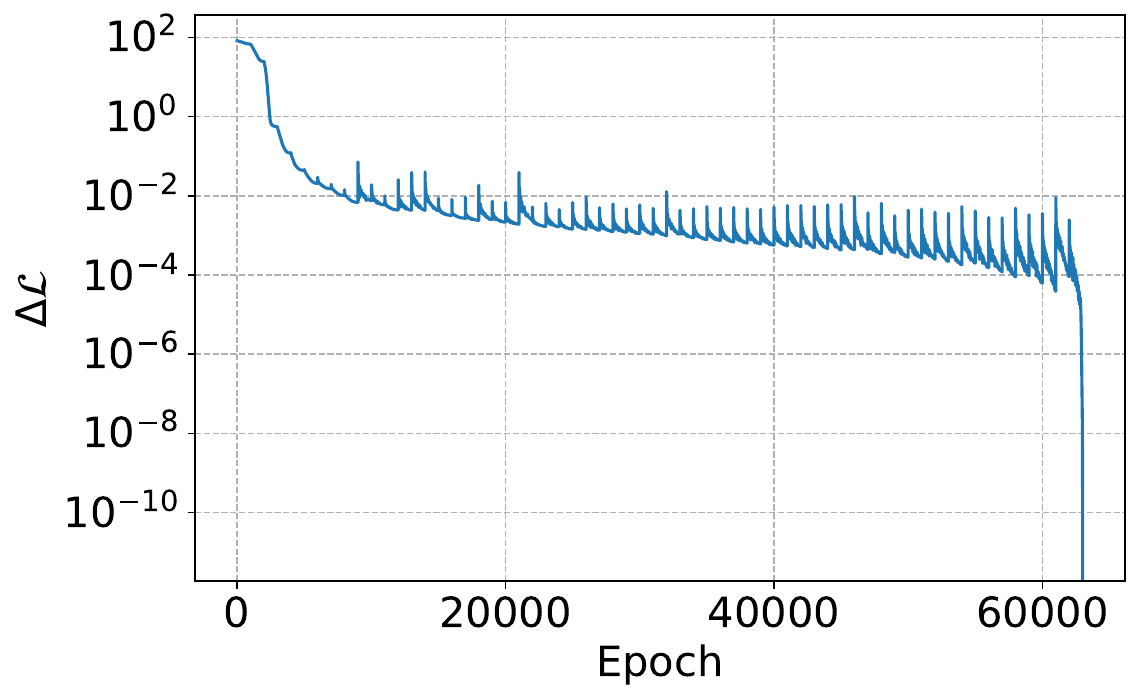} 
\caption{Example of a typical learning curve, indicating the loss function minus its minimum value (reached at convergence), $\Delta \mathcal{L}$, as a function of training steps for energy-based training of the 1D ground state for all $U$ values with $M=N=7$. 
} 
\label{fig-1d-groud-deltaloss}
\end{figure}

We choose the training set of $U$ to extend from the weakly interacting regime, starting with $U=0.01$, to the strongly repulsive regime at $U=100$. This very large range of $U$, spanning $4$ orders of magnitude, allows us to fully test the convergence and stability of \HubNet. In order not to prejudice the training of the NN to either regime, we choose $9$ training $U$ values equally spaced on a $\log$-scale,
\begin{equation}
\log_{10} U^\text{(train)} =\left\{\pm 2, \pm 1.5, \pm 1, \pm 0.5, 0 \right\}. 
\label{eq:Utrain}
\end{equation}
To make the training process converge across this wide range of $U$ values, we reduce the maximum learning rate from $0.01$, as used by \citet{Zhu2023}, to $5 \times 10^{-6}$. We also find that changing the optimizer from stochastic gradient descent to the adaptive momentum method implemented in \textsc{Adam} \cite{Kingma2017} enables the NN to yield a prediction of $E_0$ being $\approx 100$ times more accurate. 
Furthermore, we observe that the training does not converge if a constant learning rate is utilized.  Therefore, we continue to apply \HubNet's cosine-annealing scheme in the training process, resetting the learning rate to $5 \times 10^{-6}$ every $1000$ epochs. Figure \ref{fig-1d-groud-deltaloss} gives a typical learning curve of training for 1D ground state energy, i.e., minimizing $\mathcal{L}_0$. We observe that the loss function oscillates dozens of times before reaching a minimum due to the application of the cosine-annealing scheme.
In our case, convergence is reached when the standard deviation of the last $50$ iterations falls below $10^{-7}$.

Once the training is complete, we proceed to use the NN to predict energy and state features at $69$ ``out-of-data'' \cite{Arora2021} $U$ values equally spaced on log scale in the interval $\log_{10} U\in[ -2.1250, 2.1250]$, corresponding to $0.0075\leqslant U \leqslant 133$. 
Unless otherwise stated, the training and test $U$-sets will be the same for all the studies presented in the manuscript. 
Regarding the 1D ground state, we shall be interested in discussing how the NN performs in the superfluid phase, close to the transition, and in the Mott phase. Here we choose $U= 0.2054$ ($\approx 0.21$), $2.054$ ($\approx 2.05$) and $20.54$, corresponding to mid-points between adjacent training values, to be representative for the three regions. The latter $U$ values will also be taken as a reference in 2D and when analysing the excitation spectrum. 

\subsubsection{\label{sec:ml-further} Physics-based improvements}

While additional ML hyperparameter tuning---beyond learning rate---(e.g., depth/width of layers, regularization, batch size, optimizer schedules) can further improve numerical performance, we deliberately avoid a purely ML-centric optimization loop. Instead, we focus on physics-based refinements that directly influence what the network must represent, aiming at \emph{(i)} working with an apt loss function, and \emph{(ii)} selecting an optimal output activation. 
In particular, these considerations play a most significant role in the study of the system's excitation spectrum. 
As we show later, a careful choice of output activation turns out to be central: Given that we intend to have no explicit control over the bounds for the output variable $u_k$ (which acquires values in $[-17,12]$ in our 1D simulations), the function $\sigma(u)$ directly controls how well the NN can resolve very small wave-function amplitudes. In combination with a loss function that targets the many-particle state structure via the information encoded in the \revisionRAR{GDF}, we enable the NN to successfully describe the evolution of the eigenstate-structure statistics deep in the spectrum bulk as a function of the interaction strength, accounting for the emergence of a quantum chaos regime. 

Prioritizing such physics-informed choices keeps the model interpretable and ensures that observed gains reflect improved physical resolution rather than purely algorithmic tuning.

\section{\label{sec:ground-state} Ground state properties}

\subsection{\label{sec:ground-state-1d} The ground state 
for the 1D chain}

We focus on the BHH at unit density with $M=N=7$. 
This is the physically most interesting case in which the system, in the thermodynamic limit, exhibits a transition from a superfluid to a Mott insulator phase at 
$U_c\approx 3.3$ \cite{Kashurnikov1996,Ejima2011,Carrasquilla2013,Boeris2016}. Here, we consider the loss function in Eq.~\eqref{eq:loss}, and our choice for the activation funcion is $\sigma(u)=\exp(u)$, as considered by \citet{Zhu2023}.
\subsubsection{\label{sec:ground-state-1d-E0} Ground state energies \& fidelities}

\begin{figure}[tb]  
\centering
  \includegraphics[width=0.95\columnwidth]{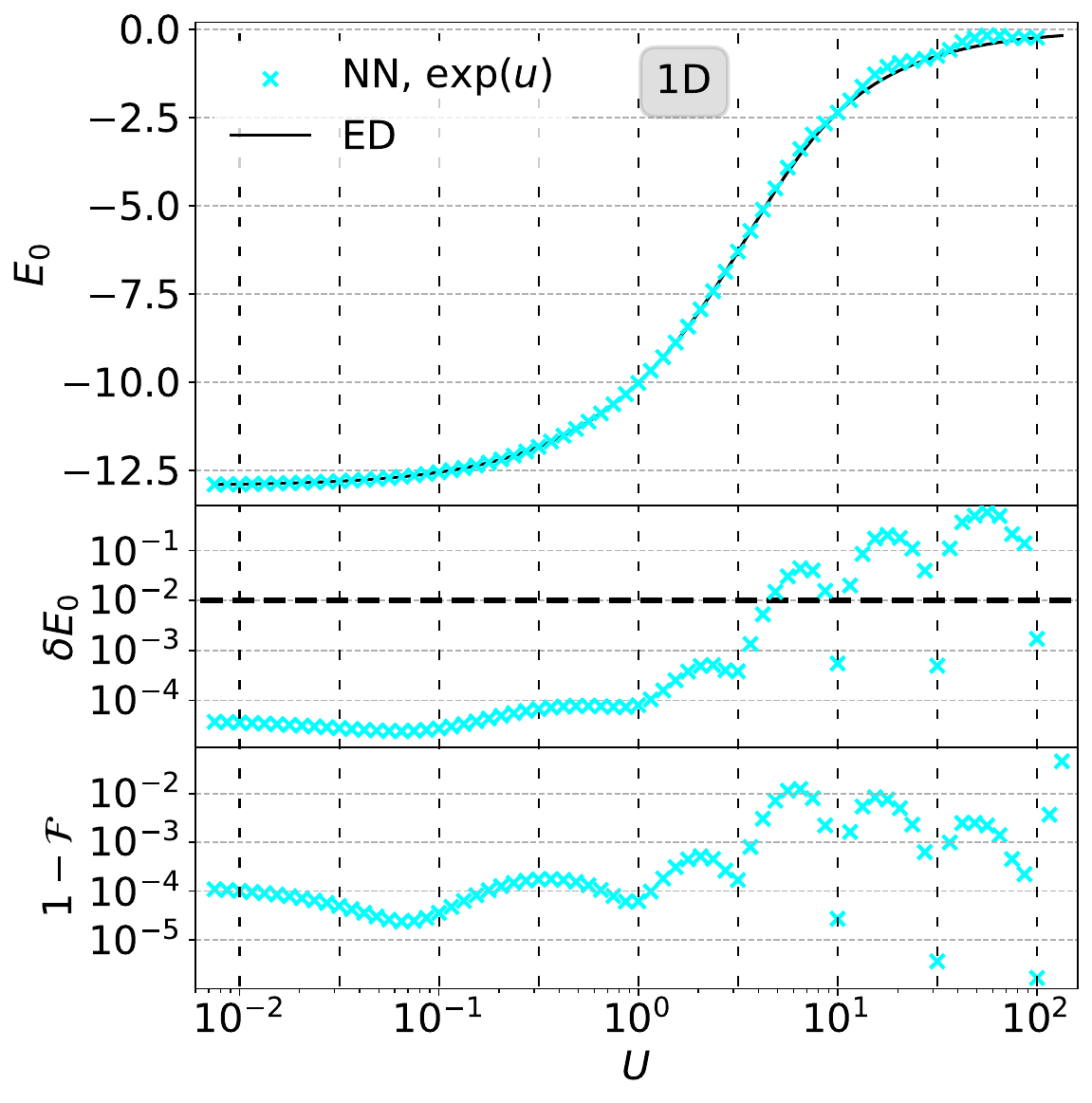} 
\caption{Accuracy of the energy-based ground state training in 1D: The upper panel shows $E_0$ for $M=N=7$ as a function of interaction  $U$ in terms of the NN (blue $\times$) and the ED (solid black line). 
The $\exp (u)$ label indicates that the exponential $\sigma$ has been used (see text).
The middle panel gives $\delta E_0$ [Eq.~\eqref{eq:relativeEkerror}] with the horizontal dashed line indicating a $10^{-2} \equiv 1\%$ value.
The infidelity $1-\mathcal{F}$ is shown in the lower panel.  
The nine vertical black dashed lines indicate the training values of $U$ [Eq.~\eqref{eq:Utrain}].
} 
\label{fig-1d-ground-state-energy-overlap}
\end{figure}

In Fig.\ \ref{fig-1d-ground-state-energy-overlap} we present the results of \HubNet\ when trained and tested as explained in Sec.~\ref{sec:ml}. 
The upper panel of Fig.\ \ref{fig-1d-ground-state-energy-overlap} compares $E_0$  obtained via NN and ED. Overall, $E_0^\text{NN}$ \revisionRAR{matches extremely} well with $E_0^\text{ED}$ across the whole range of $U$ values, up to minor deviations for $U\gtrsim 10$.
In the middle panel of Fig.\ \ref{fig-1d-ground-state-energy-overlap}, the relative error 
\begin{equation}
    \delta E_0 
    = \left|\frac{E^\text{ED}_0 - E^\text{NN}_0}{E^\text{ED}_0}\right|
    \label{eq:relativeEkerror}
\end{equation}
is plotted against $U$.
We first note that, as expected, $\delta E_0$   
for all $9$ of the training $U$ values is $< 1\%$. 
This high level of accuracy remains for test $U$ values up to $U\lesssim 5$.
As $U$ increases further, the relative deviation $\delta E_0$ becomes larger due to the fact that the exact $E_0$ progressively approaches zero (the exact ground state energy for $U=\infty$) but remains small in absolute terms.

In the lower panel of Fig.\ \ref{fig-1d-ground-state-energy-overlap}, we show the state \emph{infidelity} $1-\mathcal{F}$ as a function of $U$, where the \emph{fidelity} $\mathcal{F}$ is defined as
\begin{equation}
\mathcal{F}=\braket{\Psi_0^{\text{ED}}|\Psi_0^{\text{NN}}}.
    \label{eq:fidelity}
\end{equation}
A smaller value of infidelity corresponds to a better agreement between $\ket{\Psi_k^{\text{ED}}}$ and $\ket{\Psi_k^{\text{NN}}}$. 
We see that 
the infidelity remains below $\sim 1\%$ over the whole $U$ range except for the largest out-of-data $U$ value.
This is a remarkable result: the NN, slightly modified from the original \HubNet\ set-up in terms of improved learning rate and better ML optimizer, can now faithfully reproduce the ED results for $E_0$ and produce a good overlap with $\ket{\Psi_0}$. 
Training only for $9$ values of $U$ yields a NN configuration that can provide a full and reliable parameter sweep over four decades of the interaction strength, and that is available to users of the BHH. 

\subsubsection{Fock basis coefficients and ground-state-structure statistics} 
\label{sec:ground-state-1d-psi}
\begin{figure*}
\centering 
(a)\includegraphics[width=0.6\textwidth]{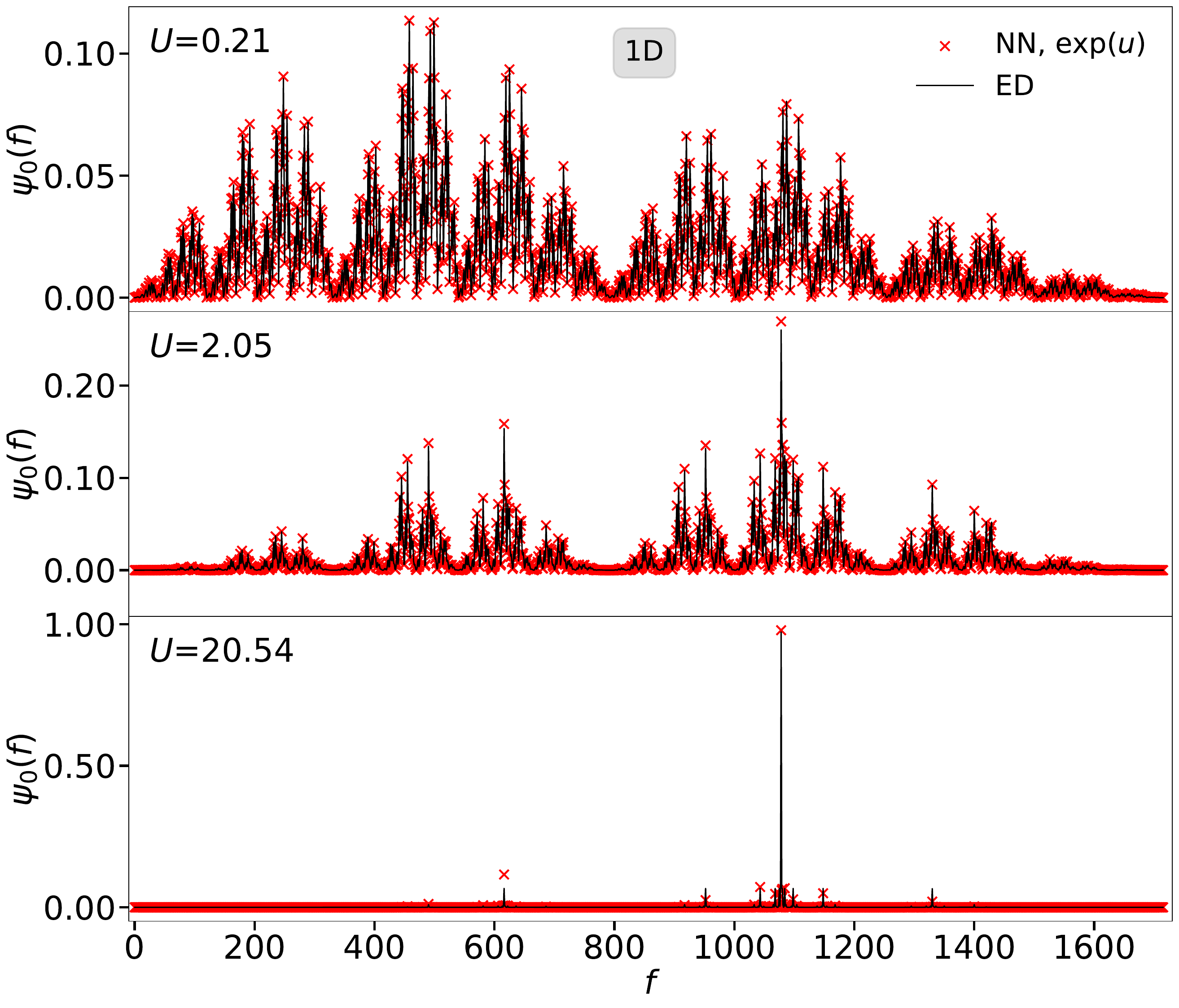} 
(b)\includegraphics[width=0.302\textwidth]{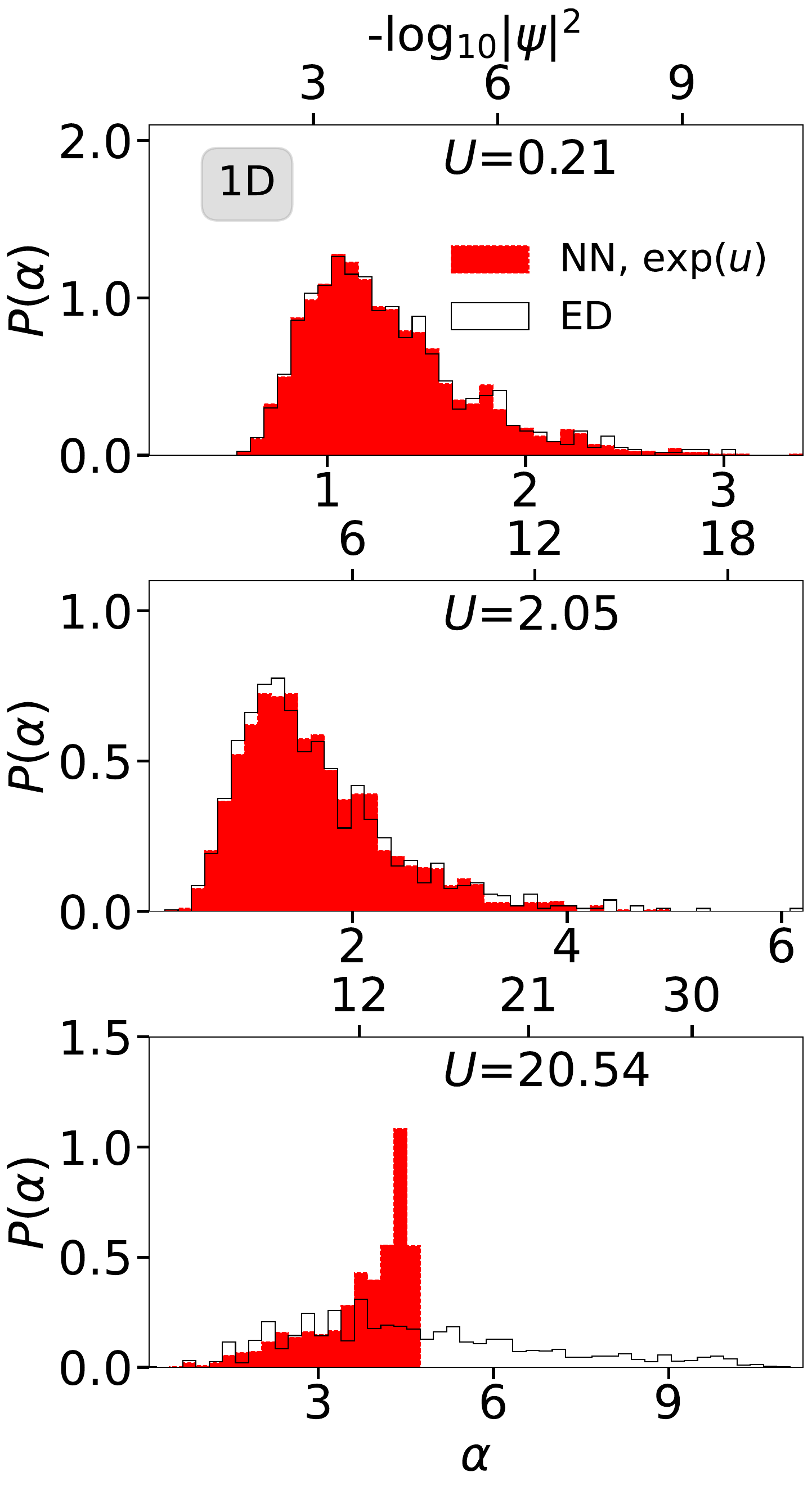} 
\caption{
Wave function properties from energy-based ground state training in 1D (cp.\ Fig.\ \ref{fig-1d-ground-state-energy-overlap}): 
(a) Coefficients $\psi_0(f)$ and (b) distribution $P(\alpha)$ obtained via the NN ($\times$, red) and the ED (black lines) at $U=0.21$ (top), $U=2.05$ (middle) and $U=20.54$ (bottom) for $M=N=7$. The horizontal Fock-space index in (a) goes from $1$ to $\text{dim}\mathcal{H}=1716$ in integer steps. Lower and upper horizontal axes in each (b) panel show the equivalent values of $\alpha$ and $|\psi|^2$, respectively.
} 
\label{fig-1d-ground-state-psi}
\end{figure*}

We now turn to investigating the agreement of individual Fock state coefficients $\psi_0(f)$ between $\ket{\Psi^\text{ED}_0}$ and $\ket{\Psi^\text{NN}_0}$.
In Fig.\ \ref{fig-1d-ground-state-psi}(a), we plot the sequence $\psi_0(f)$ as a funcion of the index $f$ at the 
values $U=0.21$, $2.05$, and $20.54$ mentioned above.
For all three $U$ values across the emerging superfluid and Mott insulator regimes---entailing a widely varying state structure, 
the amplitudes of the wave-function can be captured very well. To be specific, at $U=0.21$, the NN predicts correctly $\psi_0(f)$ for nearly each Fock index $f$. As $U$ increases to $U=2.05$, the NN can still determine the positions of the larger amplitudes, with very small deviations becoming visible. 
At $U=20.54$, the NN reproduces well the dominant contribution of the homogeneous Fock state $\ket{1, 1, \ldots, 1}$ at $f=1079$ and also identifies the larger subleading contributions from other basis states with minor fluctuations. 

A more complete picture of the predictive power of the NN concerning the state structure can be obtained by analysing the ditribution of wavefuncion intensities. 
We then construct the histograms 
${P}(\alpha)$ of \emph{singularity strengths} $\alpha$, defined as \cite{Rodriguez2009b}
\begin{equation}
    \alpha(f)\equiv -\frac{\text{ln} |\psi(f)|^2}{\text{ln}(\text{dim}\mathcal{H})} \geqslant 0.
    \label{eq:alpha}
\end{equation}
One can see from \eqref{eq:alpha} that small (large) intensities $|\psi(f)|^2$ yield large (small) $\alpha$ values: 
$|\psi|^2\to 1 \Rightarrow \alpha\to 0$, and $|\psi|^2\to 0 \Rightarrow \alpha\to \infty$.
We compare the resulting ED and NN distributions 
in Fig.\ \ref{fig-1d-ground-state-psi}(b).
For $U=0.21$ and $2.05$, we see an excellent agreement, and the NN reproduces remarkably well the state amplitudes across their entire range, up to $\alpha\approx 4$, corresponding to
$|\psi_0(f)| > 10^{-7}$. 
However, at $U=20.54$, the NN result for $P(\alpha)$, shown in the bottom panel of Fig.\ \ref{fig-1d-ground-state-psi}(b), is quite different from the ED result. 
Despite the appealing visual agreement of the amplitudes discussed earlier, the NN fails to describe correctly the set of very small wave-function intensities. In fact, $P^\text{NN}(\alpha)$ exhibits a sharp cut-off at $\alpha= 4.6$, and hence does not capture amplitudes below $4\times 10^{-8}$, whereas the exact distribution extends beyond $\alpha=9$, corresponding to $|\psi_0|<10^{-15}$.  
It must be noted that this limitation is in fact a consequence of the chosen exponential activation function: The observed range for the $u$ output of the NN in this case ranges from $-17.02$ to $0.097$. Therefore, according to Eq.~\eqref{eq:output} with $\sigma(u)=\exp(u)$,  
one can conclude that amplitudes below $4\times 10^{-8}$ cannot be resolved, as observed. 
The upper (lower) bound for $\alpha$ ($|\psi_0|$) imposed by the output activation function is then responsible for the spurious accumulation of $P^{\text{NN}}(\alpha)$ observed around $\alpha= 4$.
An appropriat\revisionRAR{ly adjusted} choice of $\sigma(u)$, however, could circumvent the limitation to sample small wave-function intensities, as we show later on. 

\subsubsection{\label{sec:ground-state-1d-training-Dq} Predicting \revisionRAR{the GDF}}

The quality of the NN prediction regarding the state structure can also be conveniently quantified by evaluating the \revisionRAR{GDF} $D_q$ 
introduced in Section \ref{sec:model-states}. 
\begin{figure}
\centering
\includegraphics[width=.95\columnwidth]{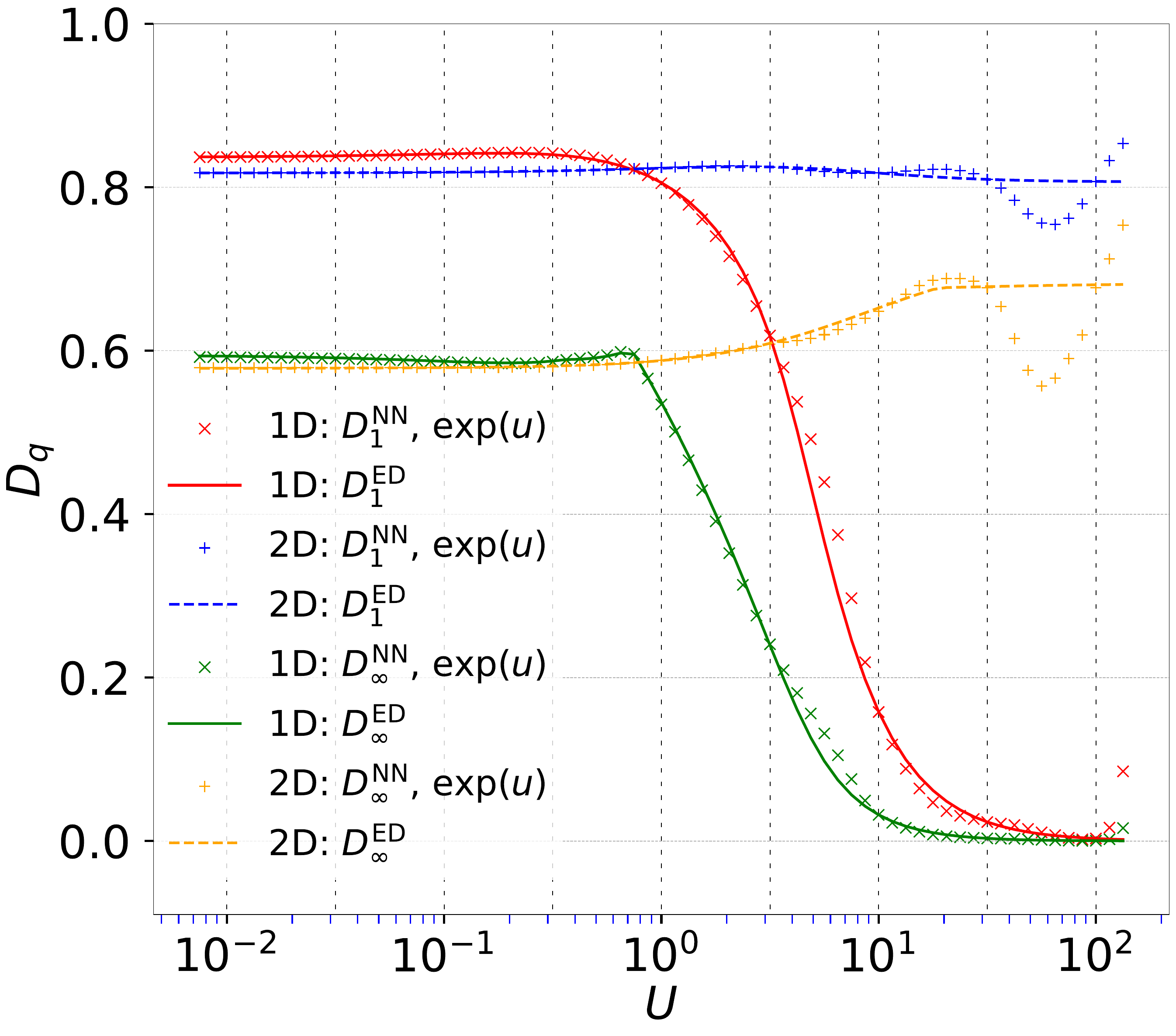} 
\caption{Estimates of $D_1$ and $D_{\infty}$ versus $U$ from the energy-based ground state training, in 1D with $M=N=7$ and 2D on a $4\times 4$ square lattice with $N=3$: 
For the 1D case, the red (green) $\times$ represent 
$D_1$ ($D_{\infty}$) obtained via the NN, whereas solid curves with the same color are the results from 
ED. In 2D, the blue (orange) $+$ symbols indicate $D_1$ ($D_{\infty}$) from the NN; solid curves of the same color show ED results.
} 
\label{fig-1d-2d-ground-state-Dq}
\end{figure}
In Fig.\ \ref{fig-1d-2d-ground-state-Dq}, we plot, e.g., the values of $D_1$ and $D_{\infty}$ against $U$ calculated from the NN and ED. 
Both $D_1^{\text{NN}}$ and $D_{\infty}^{\text{NN}}$ show a remarkable agreement with $D_1^{\text{ED}}$ and $D_{\infty}^{\text{ED}}$ across the whole range of $U$ values, and only 
small deviations become visible as the strong interacting regime is approached ($U\gtrsim 10$), in accordance with the behaviour observed for $E_0$ in Fig.~\ref{fig-1d-ground-state-energy-overlap}.
We see results of similar quality for other $D_q$ with $q= 0.5,2$ and $3$ (not shown). In particular, note how the evolution of $D_q$ from high values for weak interaction toward zero for large $U$ reveals the dramatic change in the state structure, from being large delocalized to becoming strongly localized in Fock space (manifesting the emergent superfluid to Mott insulator transition \cite{Lindinger2019}), a physically relevant development that the NN has no issues accounting for in its configuration. 
This result is very encouraging. Even when simply training to reproduce $E_0$, the 
small deviations observed in the dominant $\psi^\text{NN}_0(f)$ for $U \gtrsim U_c$ do not necessarily lead to enhanced disagreements in these $D_q$.
On the other hand, for $q<0$, we find that the predictions by the NN are considerably worse, with relative errors above $1\%$ and noticeable deviations for large interactions. This is not surprising, given that negative $q$ values rely on very small wave-function 
amplitudes, which the NN struggles to reproduce faithfully, as discussed above and observed in the bottom panel Fig.\ \ref{fig-1d-ground-state-psi}(b). 
We emphasize, nonetheless, that a solution to the latter problem is in principle at hand via an appropriate choice of the output activation function. 

\subsection{The ground state in 2D}
\label{sec:ground-state-2d}
\begin{figure}
\centering
\includegraphics[width=0.95\columnwidth]{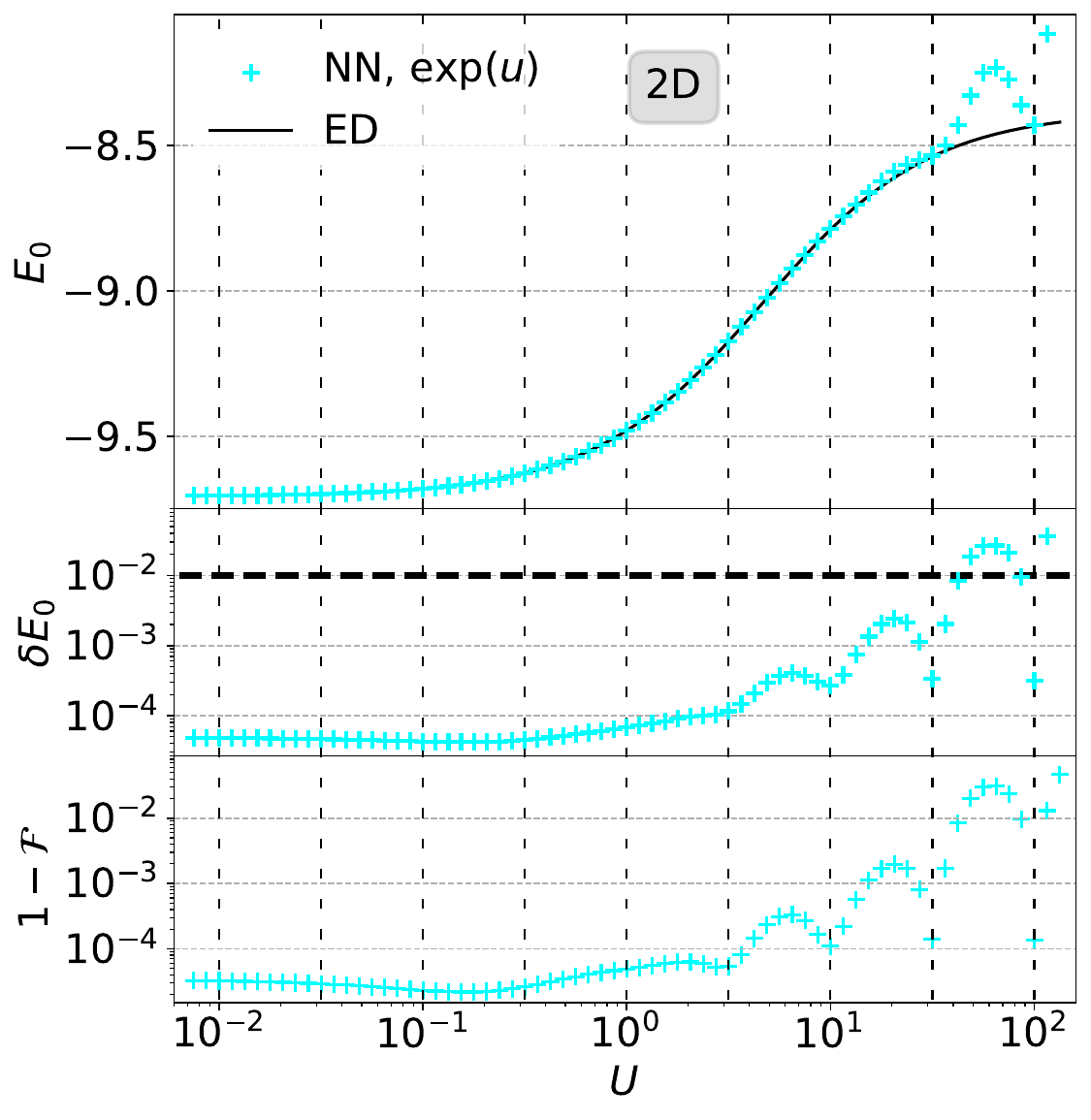} 
\caption{Accuracy of the energy-based ground state training in 2D: The upper panel shows $E_0(U)$ in terms of the NN (blue $+$) and the ED (solid black line) of a $4\times 4$ Bose-Hubbard square lattice with particle number $N=3$. 
The middle panel gives the relative error $\delta E_0$ while the infidelity $1-\mathcal{F}$ is shown in the lower panel. 
The vertical black dashed lines
and the single horizontal dashed  line 
are as in Fig.\ \ref{fig-1d-ground-state-energy-overlap}. 
} 
\label{fig-2d-ground-state-energy-overlap}
\end{figure}
\begin{figure*}
\centering
(a)\includegraphics[width=0.6\textwidth]{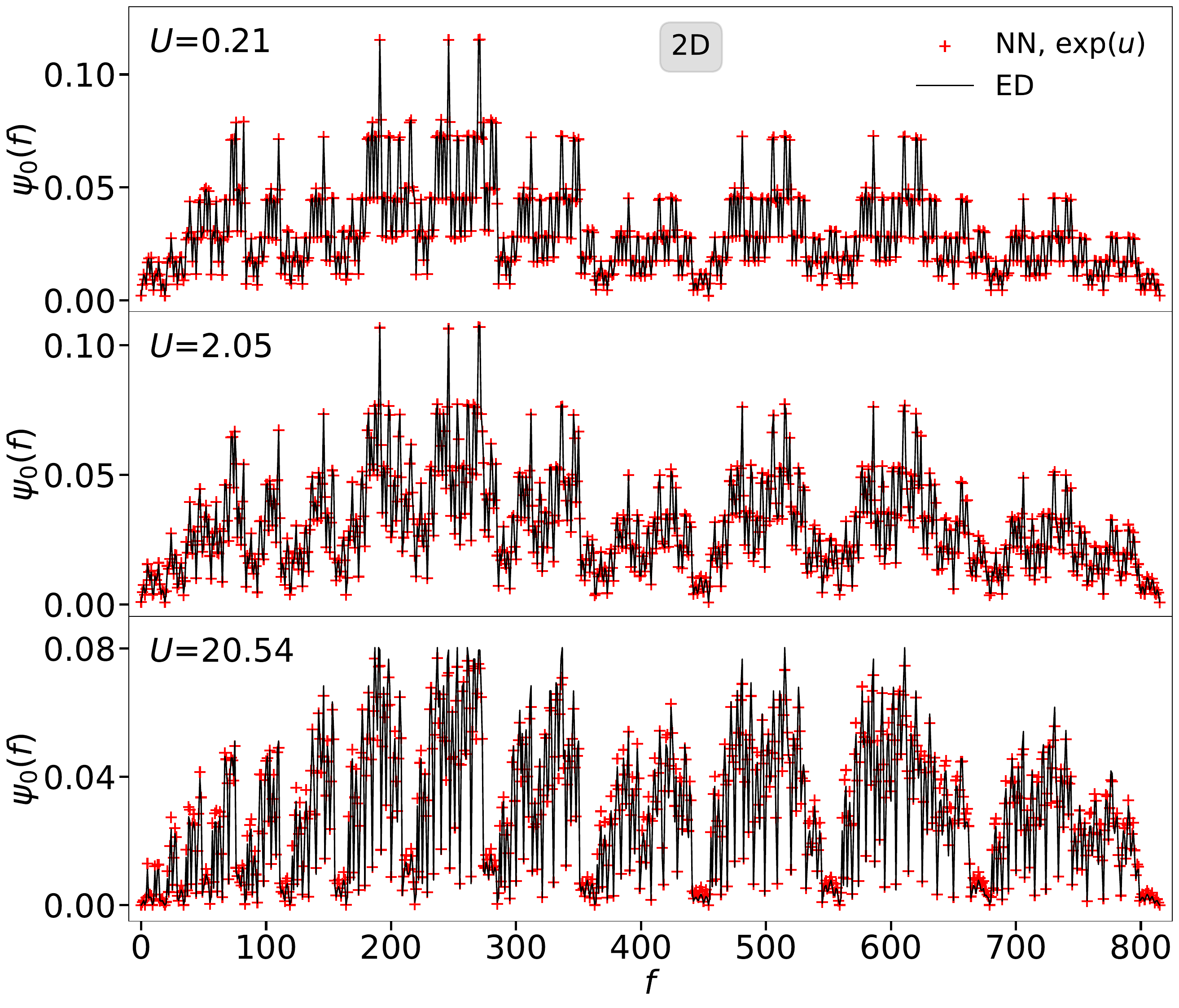} 
(b)\includegraphics[width=0.286\textwidth]{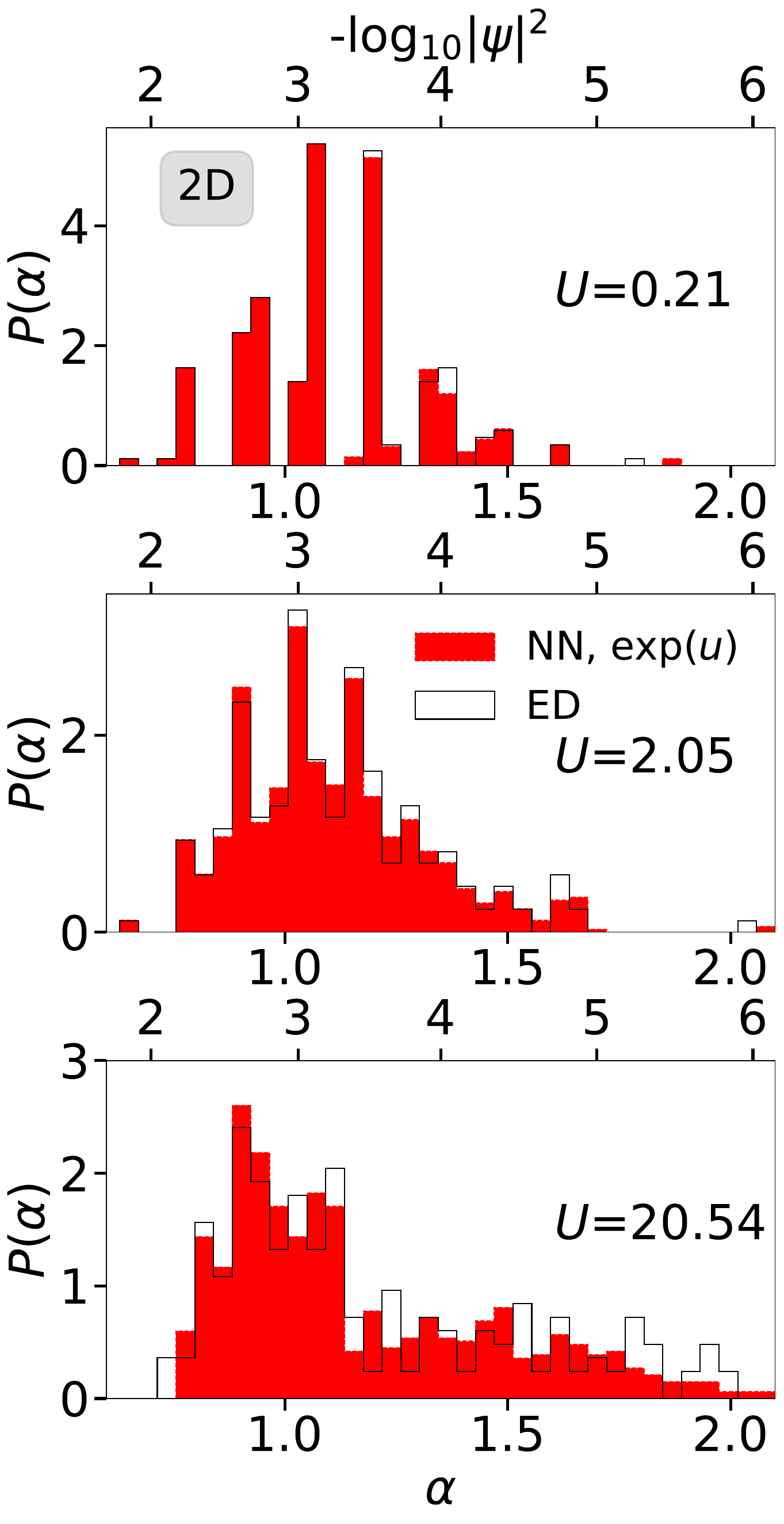}
\caption{
Wave function properties from energy-based ground state training in 2D (cp.\ Fig.\ \ref{fig-2d-ground-state-energy-overlap}): 
(a) Coefficients $\psi_0(f)$ and 
(b) distributions $P(\alpha)$ obtained via the NN ($+$, red) and ED (black lines) for a $4\times 4$ square lattice with $N=3$ ($f= 1, 2, \ldots, 816$) 
using the same three test $U$ values as in Fig.\ \ref{fig-1d-ground-state-psi}.
} 
\label{fig-2d-ground-state-psi}
\end{figure*}

Moving towards the 2D case, we 
choose system size $M=16$ in a $4\times 4$ square with particle number $N=3$, and Hilbert space dimension $\text{dim}\mathcal{H}=816$, as also considered by 
\citet{Zhu2023}. 
This gives us an opportunity to assess the performance of the NN at non-integer fillings.  Note, however, that the ground state of this system does not undergo a $U$-driven phase transition in the thermodynamic limit. As in the 1D case, we choose an energy-based training with an exponential output activation. 

We show $E_0$, $\delta E_0$ and $1-\mathcal{F}$ as obtained from the NN and ED in Fig.\ \ref{fig-2d-ground-state-energy-overlap}.
We find that the NN performs with $\delta E_0 <1\%$ up to $U\approx 50$, even better than in the 1D case. 
The prediction of the ground state, as characterized by 
$1-\mathcal{F}$, appears to be of similar quality as for 1D. 

In Fig.\ \ref{fig-2d-ground-state-psi}(a), we compare the wave-function amplitudes $\psi_0^{\text{NN}}(f)$ and $\psi_0^{\text{ED}}(f)$.  
We find an excellent match overall for the three reference $U$ values shown, and the NN only seems to underestimate some components with large amplitudes in the strong interacting regime, $U=20.54$. 
Similarly, the prediction for the distributions $P(\alpha)$ shows a very good agreement with the exact results. In particular, note that for large $U$ the prediction is much better than in the 1D case as the state does not exhibit amplitudes $|\psi_0(f)|< 10^{-3}$. Since we are not working at integer density, there is no emergent superfluid to Mott insulator phase transition, which means that the ground state structure does not change as drastically as we have seen in Fig.\ \ref{fig-1d-ground-state-psi}(a). Therefore, in this case, it should be easier for the NN to accommodate the dependence of the state features on the interaction strength. 

The NN assessment of the state structure via the evolution of the fractal dimensions 
$D_1$ and $D_{\infty}$ as functions of $U$ is shown in Fig.\ \ref{fig-1d-2d-ground-state-Dq}. The match with the exact ED results is excellent up to the value $U\approx20$, beyond which the discrepancies become more noticeable, arguably due to the cumulative effect of the small deviations observed in the dominant wave-function amplitudes for strong interactions. Nonetheless, the global tendency is correctly described, and in contradistinction to the 1D case, the lack of an emergent Mott phase in our 2D system allows the ground state to remain largely delocalized over the Fock basis, as reflected by the observed high values of $D_1$ across the entire $U$ range. 

Overall, we find a remarkable performance of the NN in 2D, similar to the 1D results, where the ground state energy and structural features can be fairly well reproduced across a wide range of interaction strength spanning four orders of magnitude. 

\section{\label{sec:excited-states} Excitation spectrum of the BHH}

In the BHH, the excited states deep in the bulk of the spectrum---those far above the ground state---are crucial for understanding the emergence of quantum chaos and the onset of ergodicity \cite{Kolovsky2004,Pausch2021}. More generally, excited states in quantum many-body systems govern the physics at finite temperatures and may obey the Eigenstate Thermalization Hypothesis (ETH) \cite{Rigol2008} suggesting that individual eigenstates can mimic thermal ensembles for local observables, bridging microscopic quantum mechanics with macroscopic thermodynamics. Moreover, phenomena like many-body localization (MBL) \cite{Yao2020, Chen2025} occur in these excited states, where, upon inclusion of disorder, thermalization may be prevented even at high energy densities---an effect with implications for quantum information storage. 
It is therefore encouraging to see that \citet{Zhu2023} have also proposed a way to construct a particular target excited state $\ket{\Psi_k}$. Their method involves an iterative construction, via the NN, of the full, orthogonalized, tower of lower-lying excited states, $\{ \ket{\Psi_l}\}$ for $l<k$.
This is obviously computationally expensive when accessing states deep in the bulk of the spectrum. Furthermore, when we attempt, e.g., the case $M=N=5$ and iteratively obtain the first $10$ excited states, we observe that sizable quantitative deviations in the $\psi_k(f)$ between ED and NN results are quickly beginning to accumulate. Unless computational power increases rapidly and well beyond current capabilities, we feel that this approach will not become practical soon.
Furthermore, while explicit knowledge of $\ket{\Psi_k}$ is 
linked to physical properties at $E_k$, one is primarily interested in these physical properties and not necessarily the details of $\ket{\Psi_k}$. In the following, we will introduce a strategy that leads us to train the same NN on average physical properties with good success.

\subsection{\texorpdfstring{$\boldsymbol{D_q}$}{$D_q$}-based training for excited states}
\label{sec:excited-states-Dq-training}
As discussed above, one needs to give up the possibility of predicting the precise Fock expansion of high excited states, as this is a daunting task. Instead, we rather focus on capturing correctly the many-body eigenstate structure as reflected by the behaviour of the average distribution of wave-function intensities, i.e., $P(\alpha)$, in the bulk of the spectrum, e.g., at scaled energy $\varepsilon=0.5$. These distributions encapsulate many physical properties of the system serving to analyze, 
i.a., the emergence of localization, multifractality, or the existence of a chaotic phase (as is the case for the BHH), hence to monitor the existence of ergodic and non-ergodic phases in the excitation spectrum. An efficient way to target the sought distributions is via the \revisionRAR{GDF} $D_q$, since these encode the features of $P(\alpha)$, as explained in Sec.~\ref{sec:model-states}.

We then propose as our loss function 
\begin{equation}
\mathcal{L}(\varepsilon)=\sum_{q^\textrm{(train)}} \sum_{U^\textrm{(train)}}|\overline{D}_{q}^{\text{ED}}(\varepsilon,U)-D_{q}^{\text{NN}}(U)|,
\label{eq:loss-Dq}
\end{equation}
where $\overline{D}_{q}^{\text{ED}}(\varepsilon,U)$ is the average fractal dimension $D_q$ over $50$ states with scaled energy closest to $\varepsilon$ (cp.\ Fig.\ \ref{fig-1d-2d-Dq} for $\varepsilon=0.5$) and $D_{q}^{\text{NN}}(U)$ is calculated from the resulting $\psi(f)= \sigma[u(f)]$ coefficients obtained via the NN. We aim at predicting the wave-function intensity distributions using a reduced set of only 5 fractal dimensions, i.e., five $q^\textrm{(train)}$ points, whose precise values will vary depending on our choice of activation function, as we discuss below. 

With the NN now trained using $\mathcal{L}(\varepsilon)$, the out-of-data $D^\text{NN}_q(U)$ for the full $69$ test $U$ values only provide a first indicator of the predictive power of the NN regarding the wave-function intensity distributions that can be quantified via the relative error
\begin{equation}
    \delta D_q 
    = \left|\frac{\overline{D}^\text{ED}_q - D^\text{NN}_q}{\overline{D}^\text{ED}_q}\right|.
    \label{eq:relativeDqerror}
\end{equation} 
We will explicitly assess the goodness of the estimated $P^\textrm{NN}(\alpha)$ by comparison against the exact $P^\textrm{ED}(\alpha)$ using the Kullback-Leibler (KL) divergence \cite{Kullback1951}, which for two discrete distributions $P$ and $Q$ is defined as 
\begin{equation}
 d_\text{KL}(P, Q)= \sum_{\alpha\in \mathcal{S}_Q} P(\alpha) \log \left[ \frac{P(\alpha)}{Q(\alpha)} \right],
 \label{eq:KLdef}
\end{equation}
where $\mathcal{S}_Q$ corresponds to the support of the broader distribution $Q$. We make this latter choice to avoid divergent values of $d_\text{KL}$ in the analysis. Equation \eqref{eq:KLdef} provides a measure of the statistical distance between the distributions, and vanishes if and only if $P$ and $Q$ are identical. We shall construct $d_\text{KL}$ at every test $U$ value, setting $Q$ to be the broader distribution of $P^\text{ED}$ and $P^\text{NN}$ at each interaction strength independently. The smaller the value of $d_\text{KL}$ as a function of $U$, the better the overall agreement between the two distributions. 

\subsection{\label{sec:excited-states-1d} Excited states of the 1D chain}
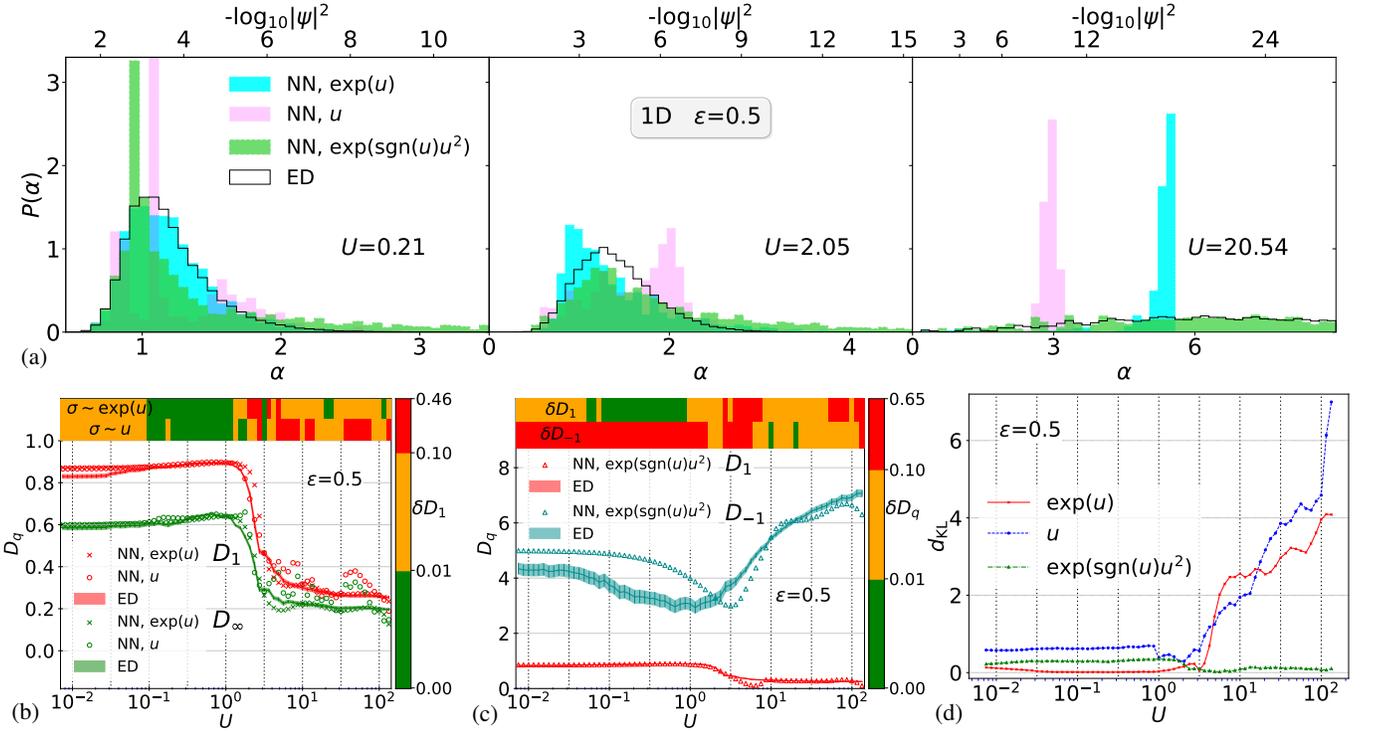
\begin{figure*}
\centering
  \begin{tikzpicture}
  \node[inner sep=0] (img) {%
    \includegraphics[width=0.98\textwidth]{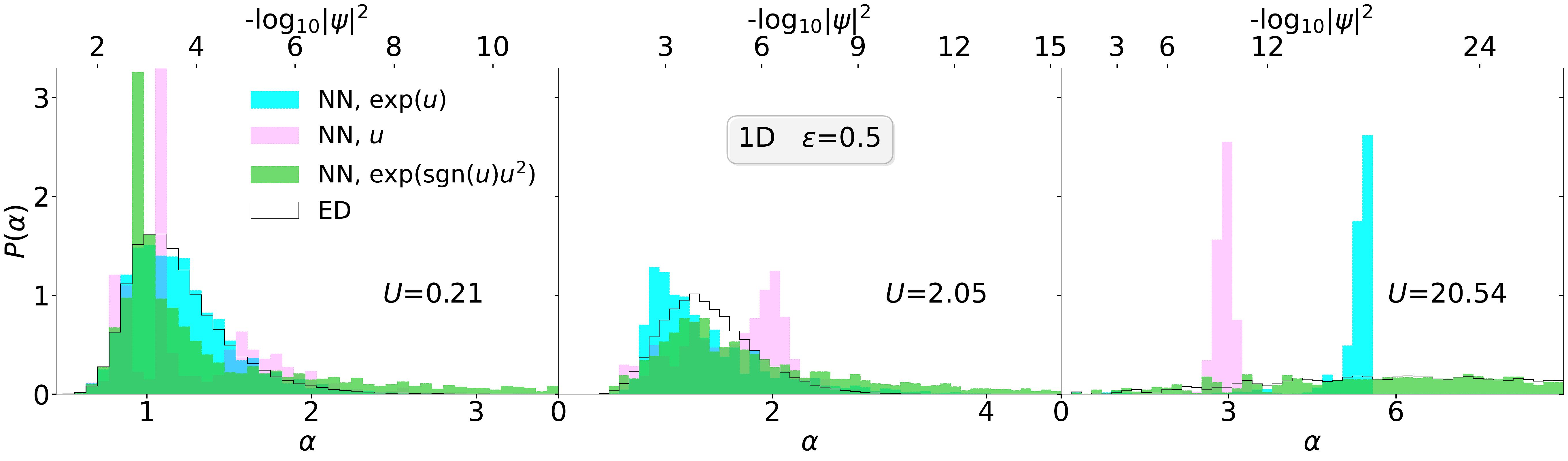}%
  };
  \node[anchor=north west] at ([xshift=-252pt,yshift=-55pt]img.center) {\small (a)};
\end{tikzpicture}
\begin{minipage}{0.335\textwidth}
\centering
\begin{tikzpicture}
  \node[inner sep=0] (img) {%
    \includegraphics[width=\textwidth]{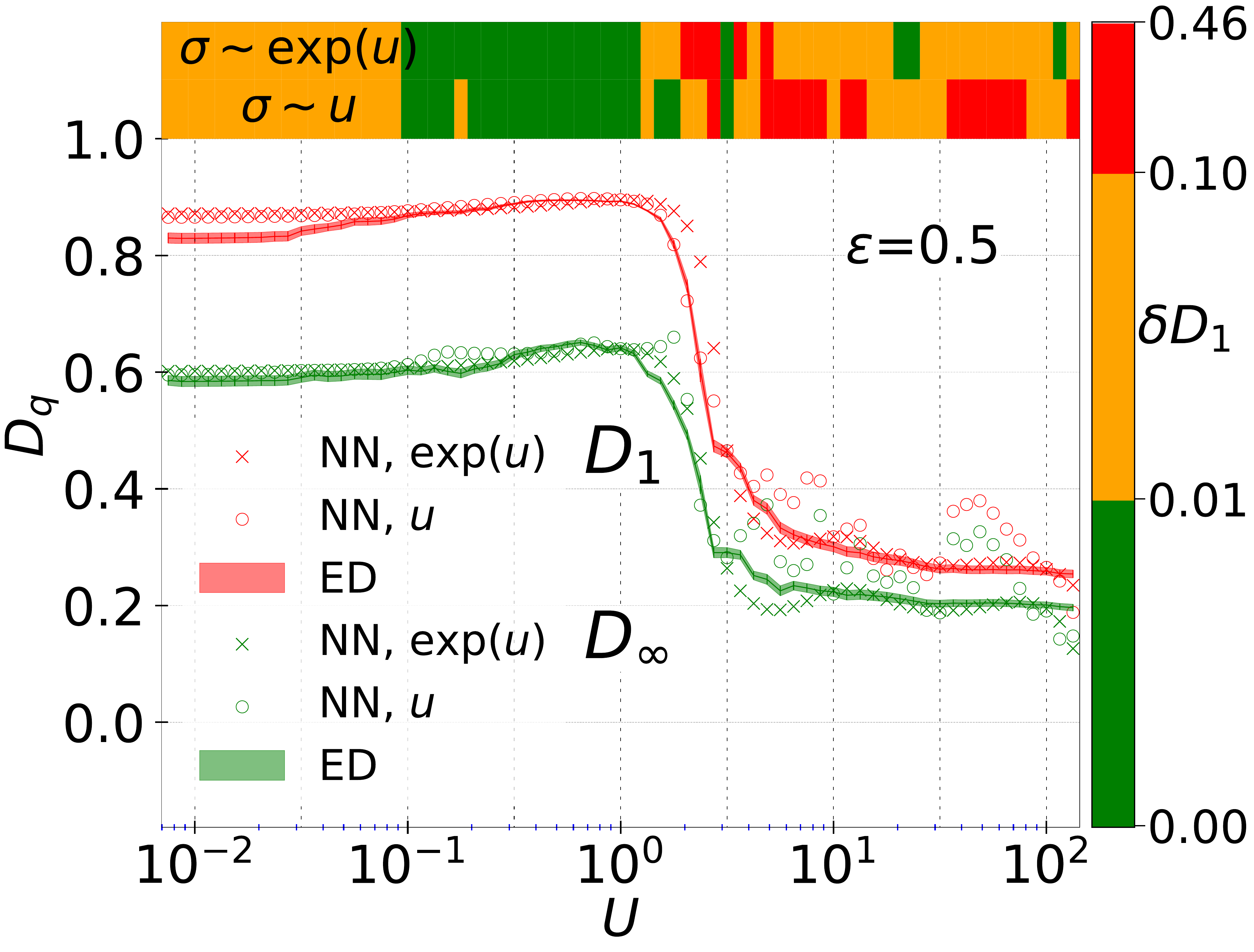}%
  };
  \node[anchor=north west] at ([xshift=-85pt,yshift=-50pt]img.center) {\small (b)};
\end{tikzpicture}
\end{minipage}\hfill
\begin{minipage}{0.335\textwidth}
\centering
\begin{tikzpicture}
  \node[inner sep=0] (img) {%
    \includegraphics[width=\textwidth]{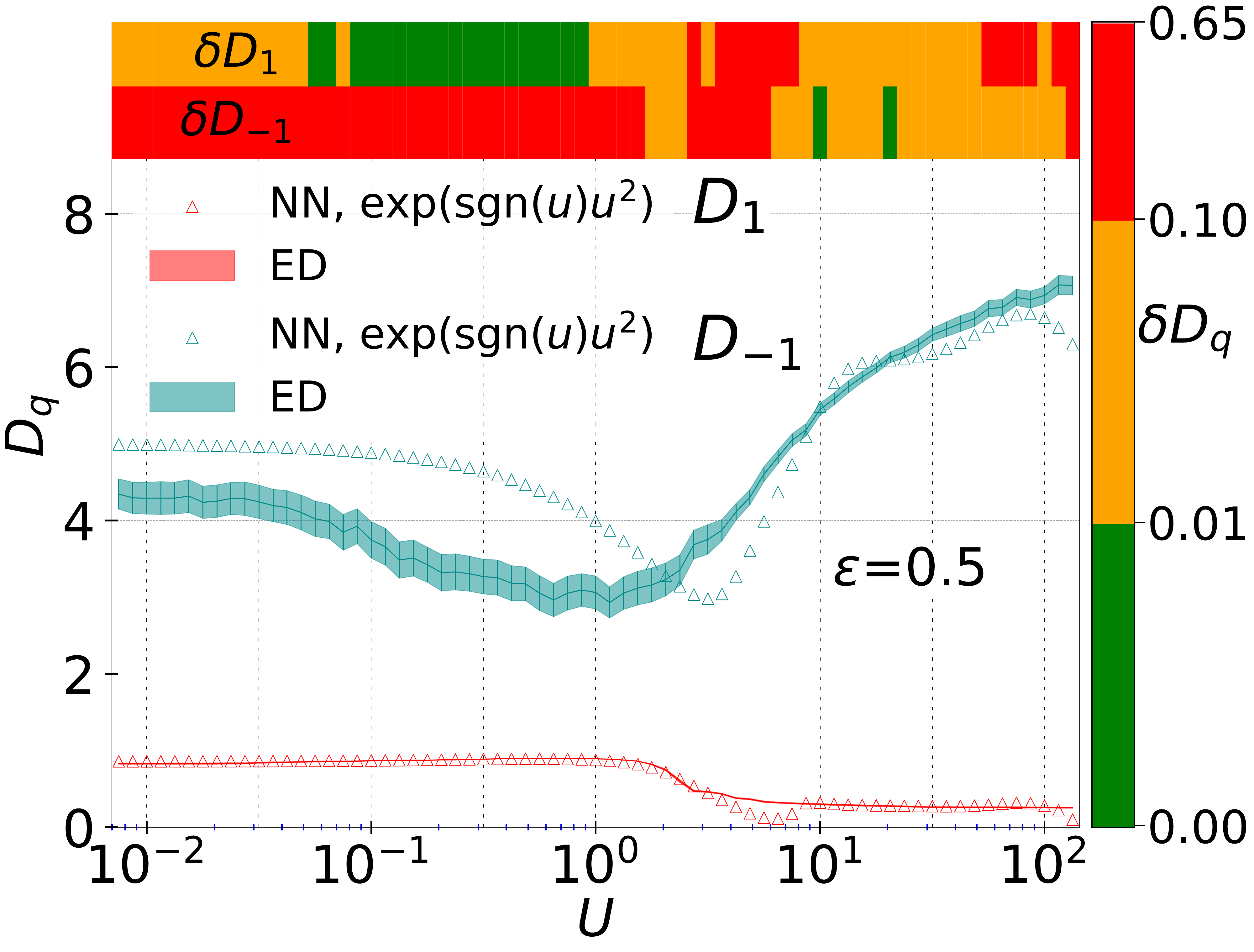}%
  };
  \node[anchor=north west] at ([xshift=-90pt,yshift=-51pt]img.center) {\small (c)};
\end{tikzpicture}
\end{minipage}\hfill
\begin{minipage}{0.315\textwidth}
\centering
\begin{tikzpicture}
  \node[inner sep=0] (img) {%
    \includegraphics[width=\textwidth]{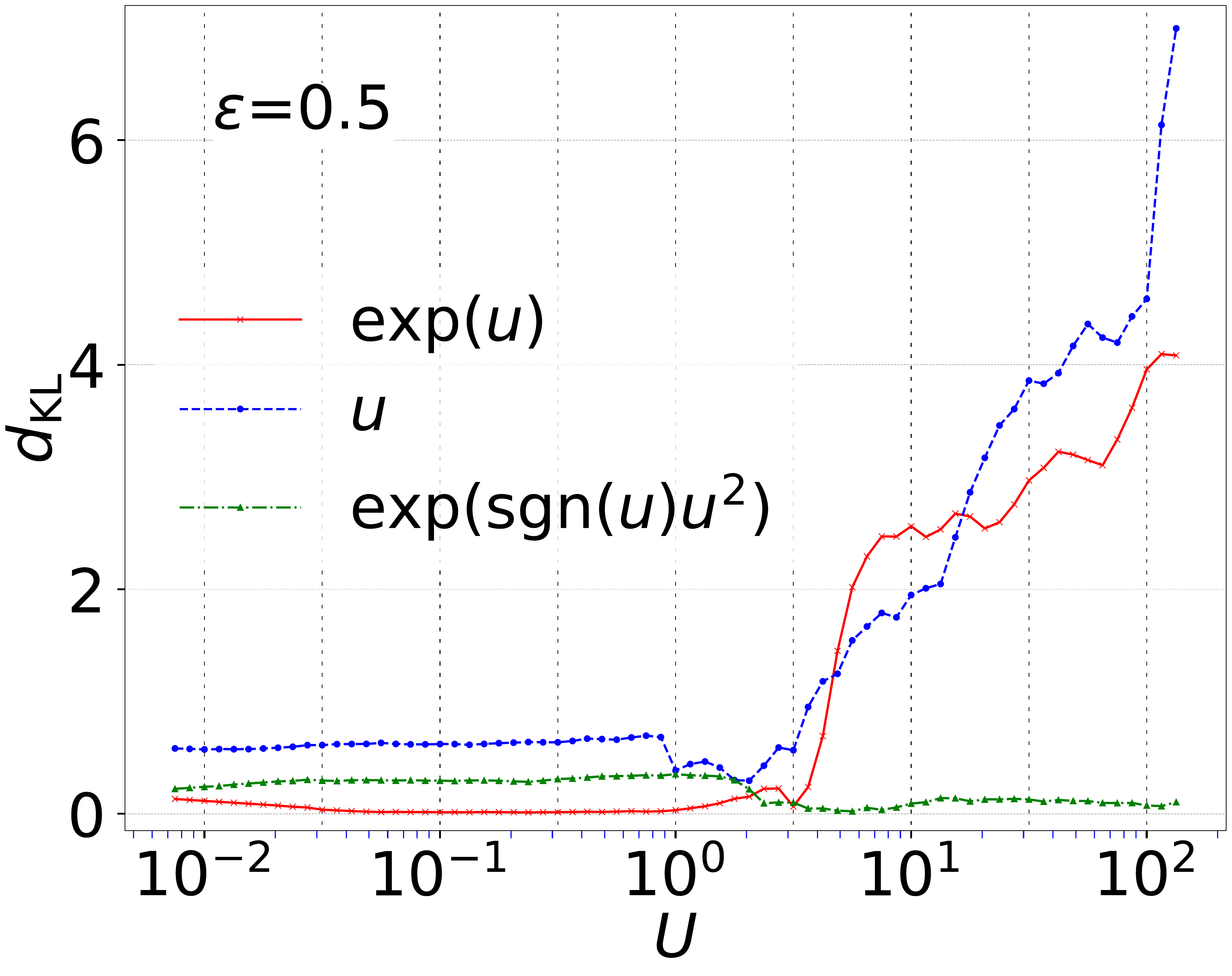}%
  };
  \node[anchor=north west] at ([xshift=-80pt,yshift=-51pt]img.center) {\small (d)};
\end{tikzpicture}
\end{minipage}
\caption{$D_q$-based excited state training in 1D 
for the 1D BHH with $M=N=7$ at $\varepsilon=0.5$ and different output activation functions. 
(a) $P(\alpha)$ at $U=0.21$ (left panel), $U=2.05$ (middle), and $U=20.54$ (right). 
The solid black line shows $P(\alpha)$ from ED. The cyan and magenta histograms 
correspond to exponential and linear $\sigma(u)$ with $q_{+}^\textrm{(train)}$, respectively, while light green follows from $\sigma(u)=\text{exp}\left[\text{sgn}(u)u^2\right]$ with $q_{-}^\textrm{(train)}$. 
Panel (b) gives $D_1$, $D_{\infty}$ 
as functions of $U$. The lines are ED results, with shadings giving the standard-error-of-mean.
Symbols represent NN results for the indicated $\sigma(u)$ 
and $q_{+}^\textrm{(train)}$.
The top color bars show $\delta D_1$ for the two $\sigma$'s, with the scale indicated by the color bar on the right. 
Vertical dashed lines mark $U^\textrm{(train)}$ values. 
Panel (c) gives $D_1(U)$, $D_{-1}(U)$ 
based on $q_{-}^\textrm{(train)}$. Colors and symbols are analogue to panel (b).
Panel (d) displays $d_\text{KL}(P^\textrm{NN},P^\textrm{ED})$ [Eq.~\eqref{eq:KLdef}] versus $U$ for the three activation functions considered. 
} 
\label{fig-1d-excited-states-alpha-Dq-KL-05-all}
\end{figure*}
\begin{figure*}
\centering
\begin{tikzpicture}
  \node[inner sep=0] (img) {%
    \includegraphics[width=0.98\textwidth]{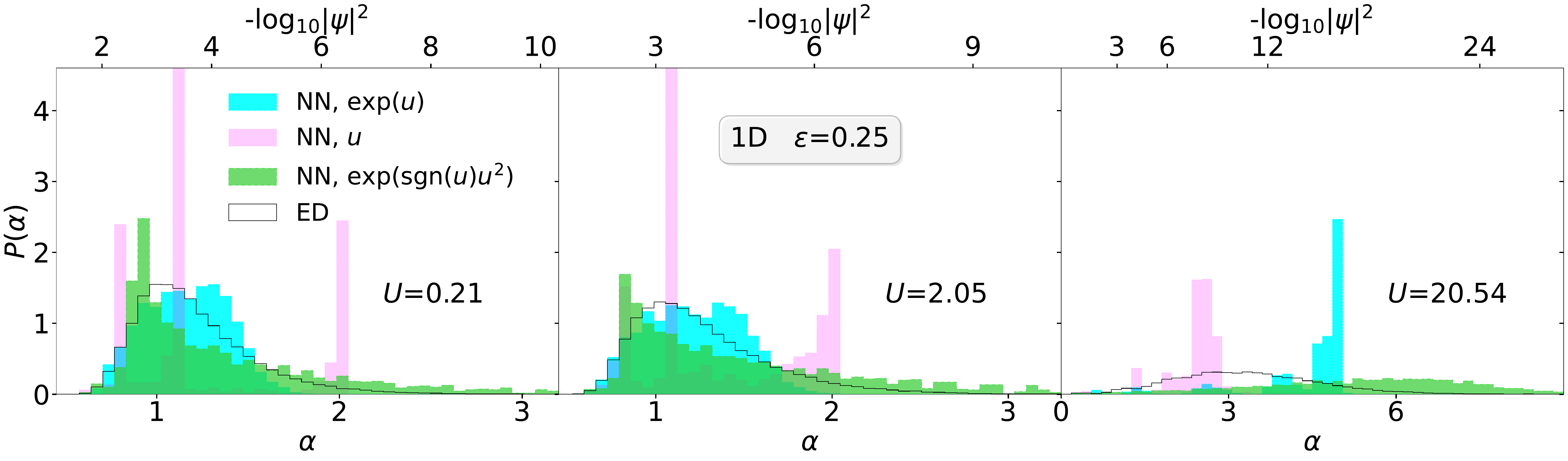}%
  };
  \node[anchor=north west] at ([xshift=-252pt,yshift=-55pt]img.center) {\small (a)};
\end{tikzpicture}
\begin{minipage}{0.335\textwidth}
\centering
\begin{tikzpicture}
  \node[inner sep=0] (img) {%
    \includegraphics[width=\textwidth]{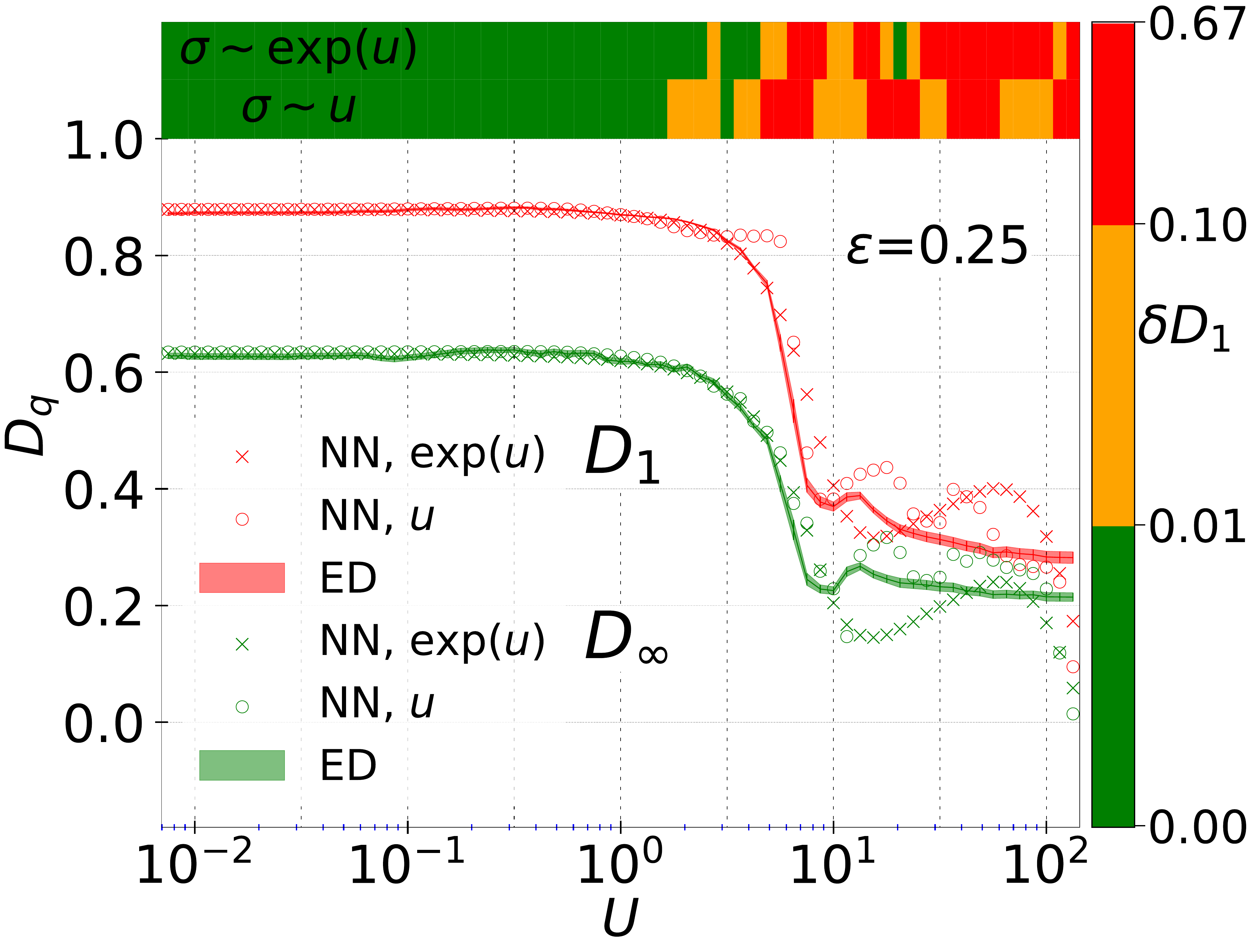}%
  };
  \node[anchor=north west] at ([xshift=-85pt,yshift=-50pt]img.center) {\small (b)};
\end{tikzpicture}
\end{minipage}\hfill
\begin{minipage}{0.335\textwidth}
\centering
\begin{tikzpicture}
  \node[inner sep=0] (img) {%
    \includegraphics[width=\textwidth]{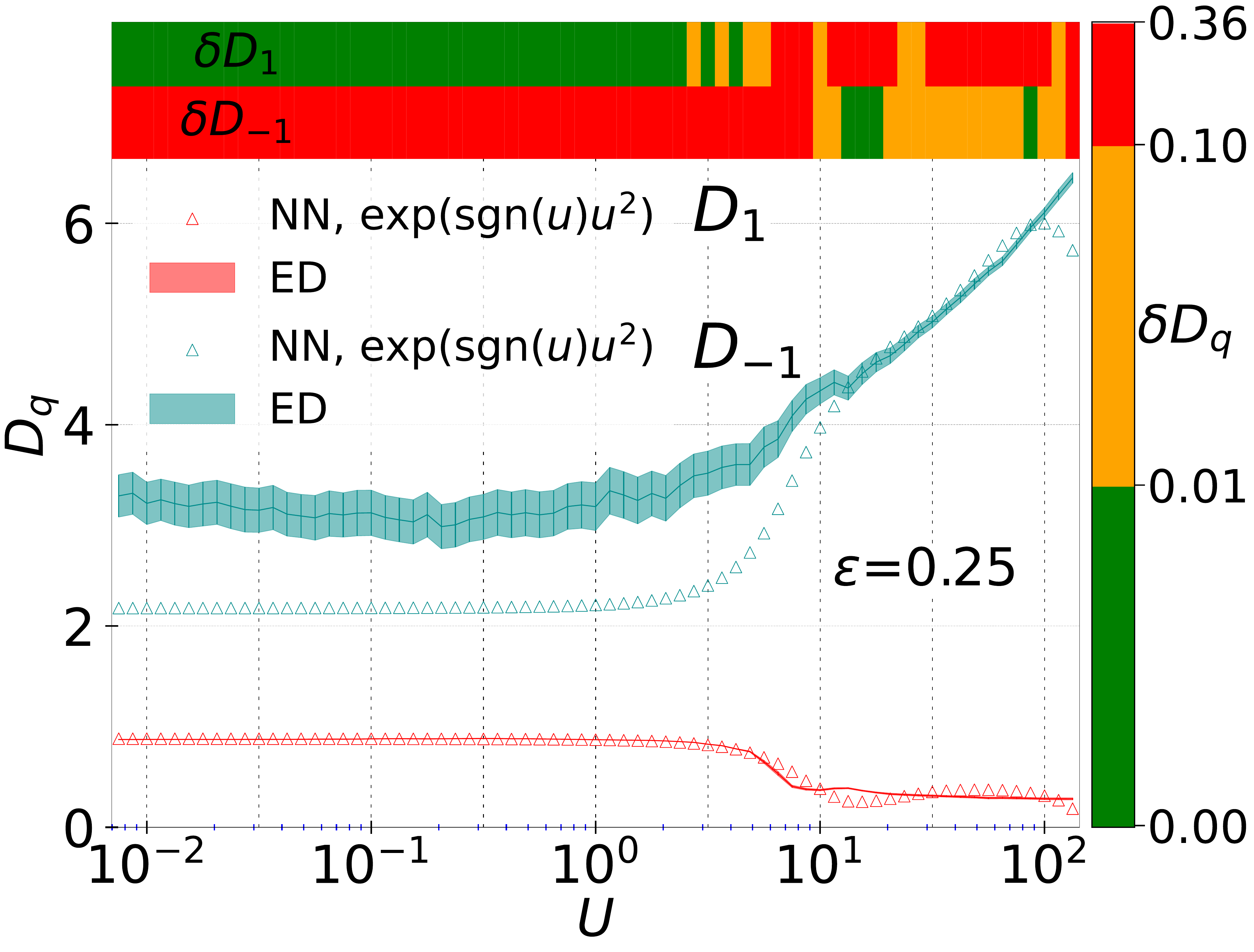}%
  };
  \node[anchor=north west] at ([xshift=-90pt,yshift=-51pt]img.center) {\small (c)};
\end{tikzpicture}
\end{minipage}\hfill
\begin{minipage}{0.315\textwidth}
\centering
\begin{tikzpicture}
  \node[inner sep=0] (img) {%
    \includegraphics[width=\textwidth]{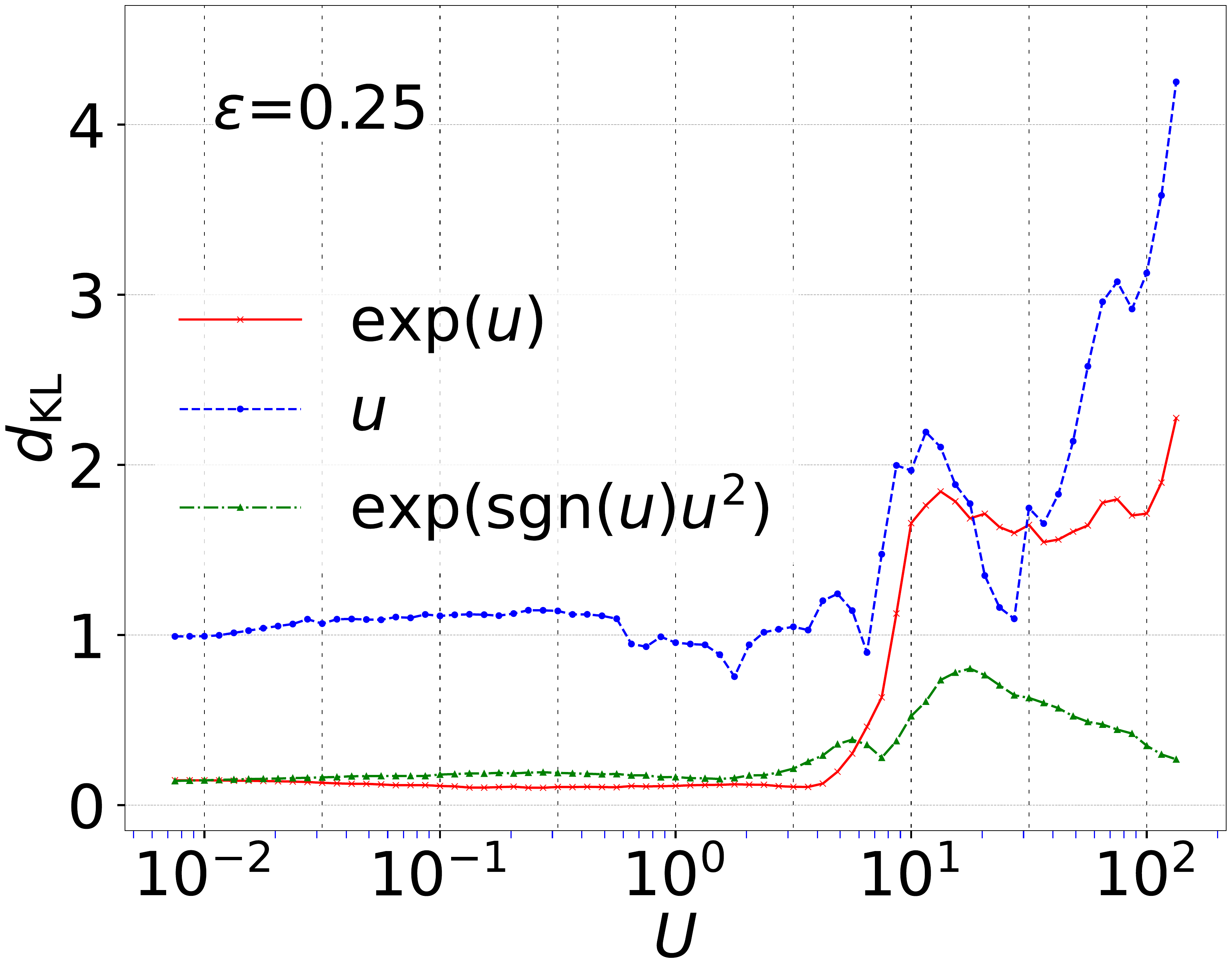}%
  };
  \node[anchor=north west] at ([xshift=-80pt,yshift=-51pt]img.center) {\small (d)};
\end{tikzpicture}
\end{minipage}
\caption{Same $D_q$-based excited state training as in Fig.\ \ref{fig-1d-excited-states-alpha-Dq-KL-05-all} except for changing $\varepsilon$ to $0.25$. The colors of histrograms 
and lines in (a), the notations of $D_{q}^{\text{NN}}$ and $D_{q}^{\text{ED}}$ in (b) and (c), the symbols and lines in (d) stay the same as in Fig.\ \ref{fig-1d-excited-states-alpha-Dq-KL-05-all}.
} 
\label{fig-1d-excited-states-alpha-Dq-KL-025-all}
\end{figure*}

We first need to decide on a choice of activation function, since, as we have seen earlier, this may affect the NN performance crucially. Note that, unlike the ground state, excited states of the BHH are no longer restricted to non-negative Fock coefficients. Therefore, one may propose a linear activation $\sigma(u)=u$, as considered by \citet{Zhu2023}. However, as we are training the NN on \revisionRAR{the GDF}, and these are only determined by the intensities $|\psi(f)|^2$ [see Eq.~\eqref{eq:Dq}], there is no need to describe the wave-function amplitude sign, and we may keep the exponential output activation $\sigma(u)=\exp(u)$. We will begin by analysing the NN results using both activation functions and the following $q$ training set, 
\begin{equation}
  q_{+}^\textrm{(train)}=\{1/2,1,2,3,\infty\},
  \label{eq:qtrainplus}
\end{equation}
in a system with $M=N=7$.  

In Fig.~\ref{fig-1d-excited-states-alpha-Dq-KL-05-all}(b), we show the predicted evolution of the \revisionRAR{GDF} $D_1$ and $D_\infty$ as functions of $U$ after the training. 
We find that the NN captures correctly the overall evolution of the GFD in the bulk of the spectrum ($\varepsilon=0.5$) across the four orders of magnitude in $U$, including the sharp decrease around $U\approx 2$ indicating the end of the chaotic phase, signalled by a pronounced increase in the localization of the excited states (hence, suppressed $D_q$ values) [cp.~Fig.~\ref{fig-1d-2d-Dq}]. The performance is better within the chaotic phase, where $\delta D_q$ is well below $10\%$, and degrades slightly for $U>2$, though the agreement with the ED results in this region is clearly better when using the exponential activation function.

The focus, however, must be put on the capability of the NN to reconstruct the wave-function intensity distributions that we show in Fig.~\ref{fig-1d-excited-states-alpha-Dq-KL-05-all}(a) for $U=0.21, 2.05$ and $20.54$.  For $\sigma(u)=\exp(u)$ (cyan histograms), we see an excellent description of the exact distribution for the lowest interaction, and a reasonable overlap at $U=2.05$. The very good performance of the NN for weak interaction strengths up to $U\approx 1$ is confirmed by the very low values of $d_\text{KL}$ observed in panel (d). In this very same $U$ range, the NN trained with the linear activation function performs noticeably worse, as indicated by the larger $d_\text{KL}$ values in (d), and visually evident in (a) (magenta histograms). Most importantly, the NN fails very significantly to describe correctly the high interaction regime, $U>3$, for either $\sigma(u)$, manifested by the obvious deviations of $P^\textrm{NN}(\alpha)$ from the ED result (black line) for $U=20.54$, and the marked surge in the Kullback-Leibler divergence in panel (d). 

The poor agreement of the distributions for large $U$ could be partially expected, since the training only included GFD for positive $q$ values, which mainly encode the behaviour of the larger wave-function intensities. A correct description of the many-body eigenstate structure for strong interactions hinges on tracking the behaviour of the small state intensities. Then, on the one hand, we need to feed such information at the training stage by considering also GFD with negative $q$, and secondly, one must overcome the limitations of the prior activation functions to yield small Fock coefficients, as discussed for the ground state in Sec.~\ref{sec:ground-state-1d-psi}. Hence, to increase the coverage of wave-function amplitudes, while maintaining the essential monotonicity of the $u\leftrightarrow \psi(f)$ mapping, we propose to use the activation function 
\begin{equation}
 \sigma(u)=\exp[ \text{sgn}(u) u^2]
 \label{eq:sigmau2}
\end{equation}
in combination with the training set 
\begin{equation}
  q_{-}^\textrm{(train)}=\{-2,-1,1,2,\infty\}.
  \label{eq:qtrainminus}
\end{equation}
The ensuing prediction for $P(\alpha)$ of this approach is shown by the green histograms in Fig.~\ref{fig-1d-excited-states-alpha-Dq-KL-05-all}(a), and the green points for $d_\text{KL}$ in panel (d). The NN captures remarkably well the distribution at $U=20.54$, including wave-function amplitudes down to $10^{-15}$. The performance of the NN configuration in this case is excellent for $U>3$ as indicate\revisionRAR{d} by the near-zero values of $d_\text{KL}$, while maintaining a very reasonable output also for low interactions. Overall, the latter training strategy yields the most well behaved and balanced prediction for the eigenstate structure of excited states across the whole interaction range. For completion, panel (c) shows the corresponding prediction for $D_1$ and $D_{-1}$ versus $U$. 

To determine if our approach remains valid at other spectral regions, we repeat the analysis for scaled energy $\varepsilon=0.25$. The results obtained for the three activation functions are presented in Fig.~\ref{fig-1d-excited-states-alpha-Dq-KL-025-all}. The conclusions are qualitatively the same as for the previous energy. The NN reproduces correctly the overall dependence of the GFD on interaction strength, capturing the pronounced change around $U\approx 5$ associated with the border of the chaotic regime. The agreement with ED is very good within the chaotic phase for positive $q$ while larger relative errors $\delta D_q$ emerge for strong interactions. When it comes to the $P(\alpha)$ distributions, the linear activation function performs very poorly for any $U$, as observed in the (a) and (d) panels, and while $\sigma(u)=\exp(u)$ works fairly well for weak and moderate interactions, it clearly fails for $U>5$. As before, the training based on Eq.~\eqref{eq:sigmau2} and $q_{-}^\textrm{(train)}$ produces the NN with the best performance throughout the entire $U$ range, albeit its accuracy seems arguably lower than for $\varepsilon=0.5$.

We have thus demonstrated that a clever $D_q$-based training combined with an appropriate choice of activation function leads to a NN configuration capable of describing correctly the physically relevant changes in the many-body eigenstate structure deep in the bulk of the spectrum across a substantial interaction range, without the need to target the precise Fock space expansion of selected excited states. 
Our activation function in Eq.~\eqref{eq:sigmau2} is \revisionRAR{guided by physical intuition}
but there might be even more efficient choices. 

\subsection{\label{sec:excited-states-2d} Excited states for the 2D case}
\begin{figure*}
\centering
\begin{tikzpicture}
  \node[inner sep=0] (img) {%
    \includegraphics[width=0.98\textwidth]{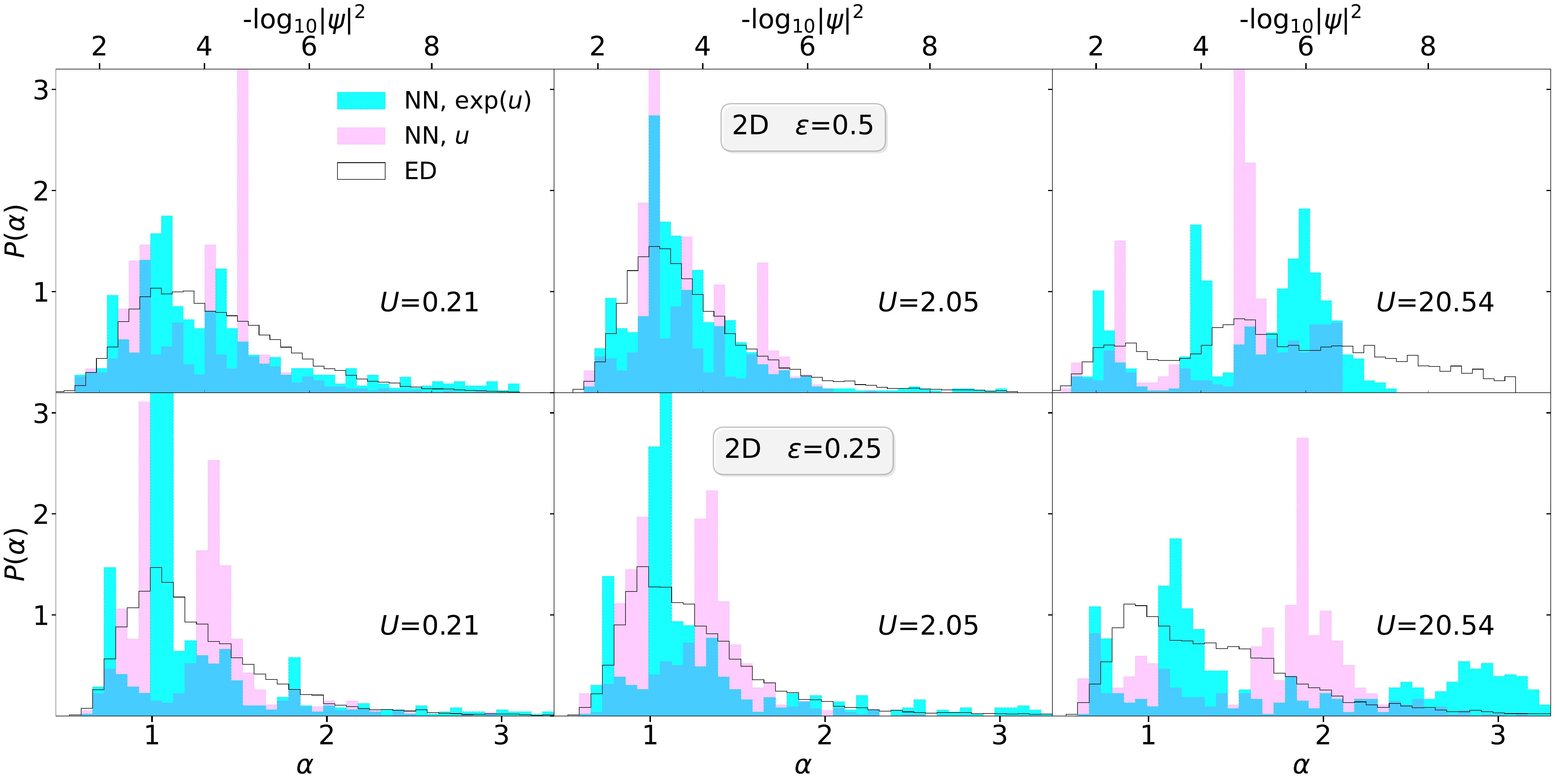}%
  };
  \node[anchor=north west] at ([xshift=-252pt,yshift=10pt]img.center) {\small (a)};
  \node[anchor=north west] at ([xshift=-252pt,yshift=-95pt]img.center) {\small (b)};
\end{tikzpicture}
\begin{minipage}{0.335\textwidth}
\centering
\begin{tikzpicture}
  \node[inner sep=0] (img) {%
    \includegraphics[width=\textwidth]{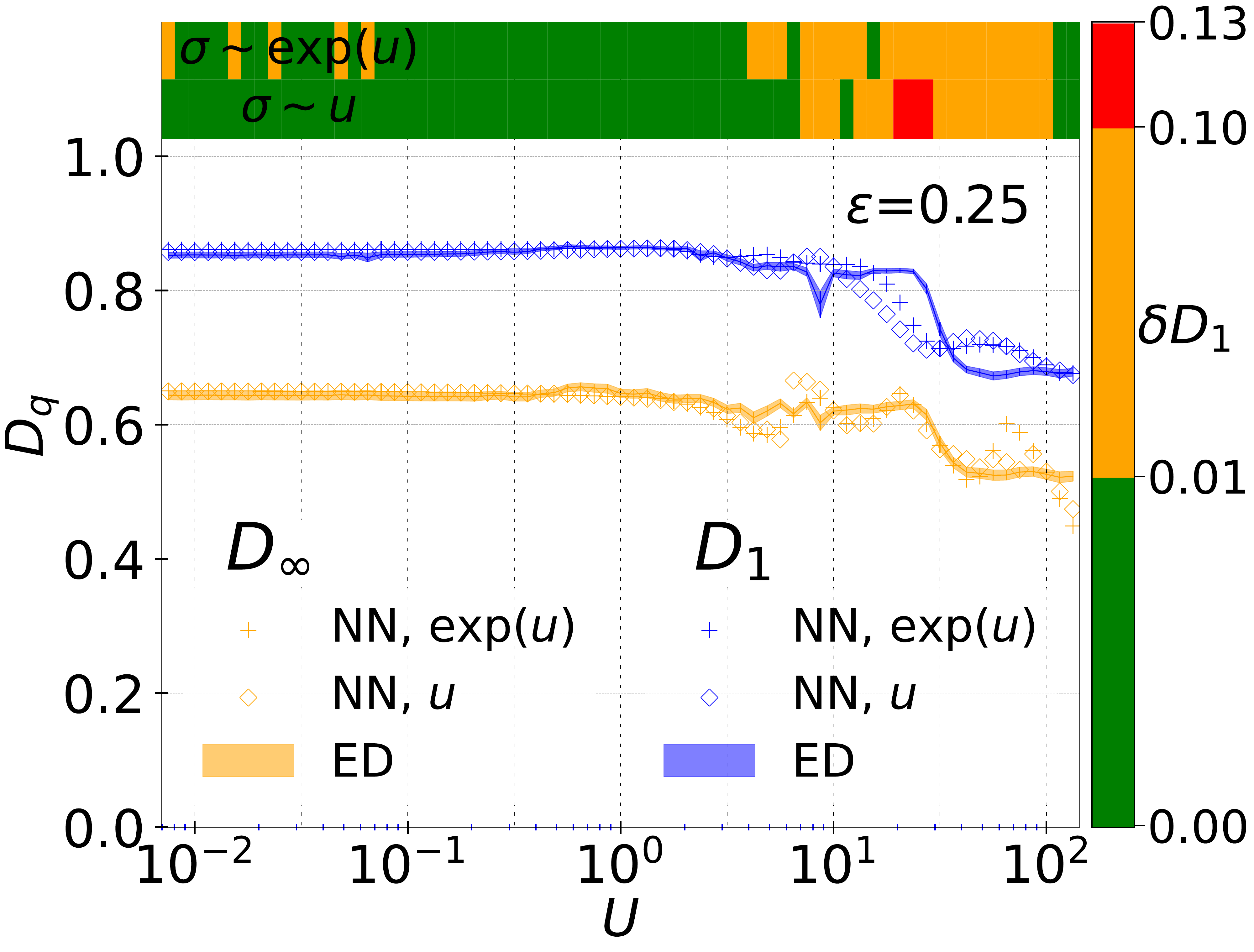}%
  };
  \node[anchor=north west] at ([xshift=-88pt,yshift=-51pt]img.center) {\small (c)};
\end{tikzpicture}
\end{minipage}\hfill
\begin{minipage}{0.335\textwidth}
\centering
\begin{tikzpicture}
  \node[inner sep=0] (img) {%
    \includegraphics[width=\textwidth]{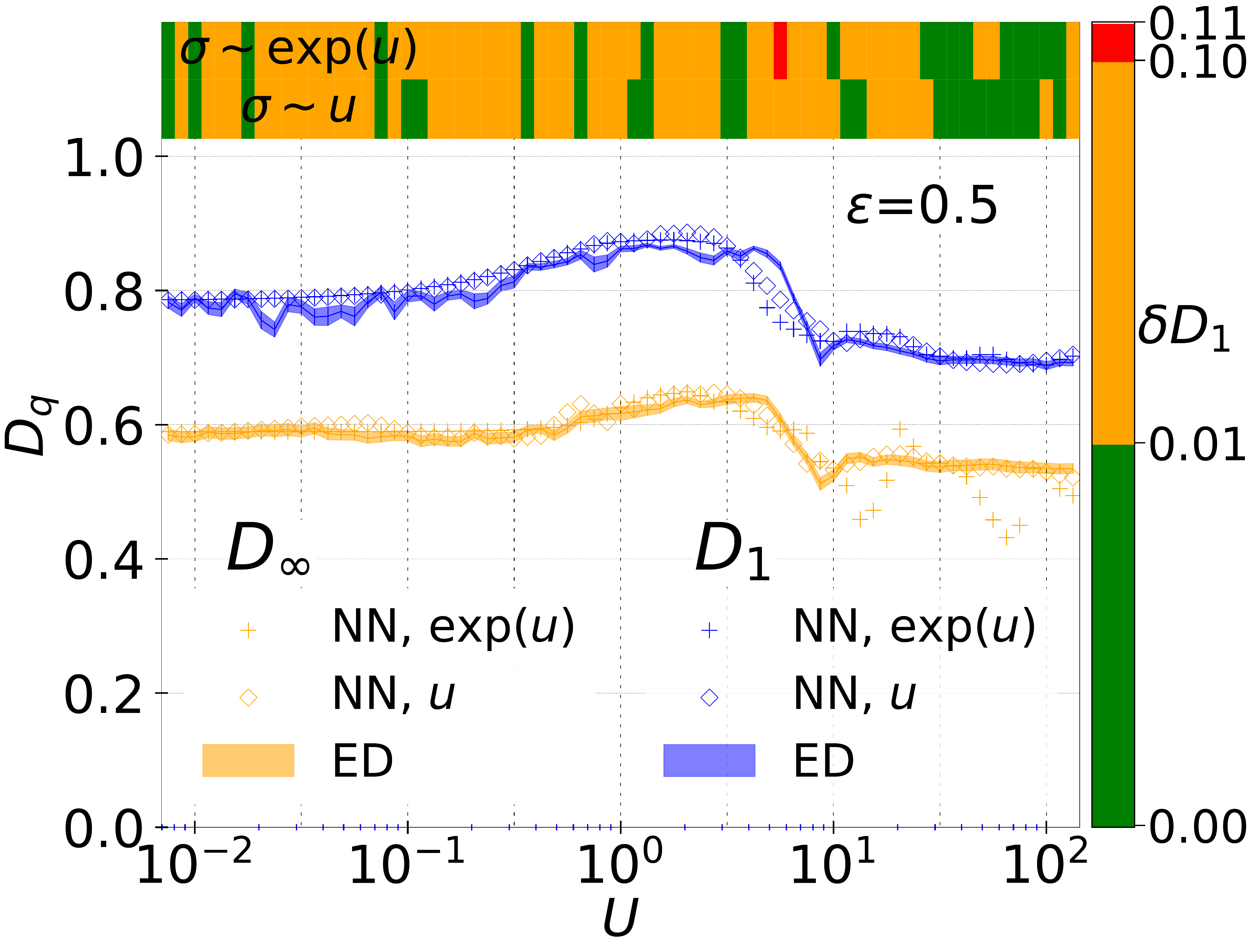}%
  };
  \node[anchor=north west] at ([xshift=-88pt,yshift=-51pt]img.center) {\small (d)};
\end{tikzpicture}
\end{minipage}\hfill
\begin{minipage}{0.315\textwidth}
\centering
\begin{tikzpicture}
  \node[inner sep=0] (img) {%
    \includegraphics[width=\textwidth]{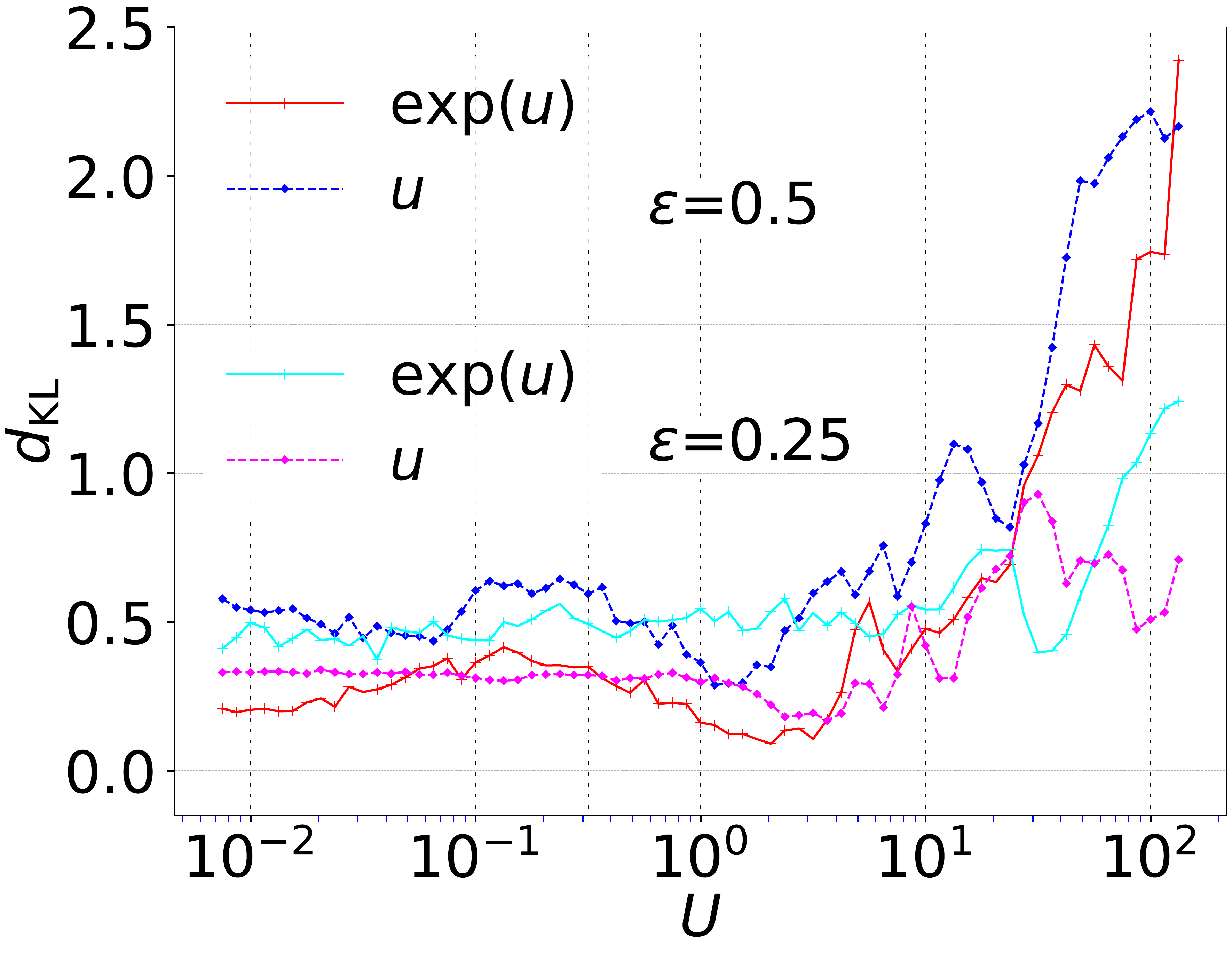}%
  };
  \node[anchor=north west] at ([xshift=-80pt,yshift=-51pt]img.center) {\small (e)};
\end{tikzpicture}
\end{minipage}
\caption{$D_q$-based excited state training for the 2D BHH on a $4\times 4$ square lattice with $N=3$, at different scaled energies, and for linear and exponential activation functions with $q_+^\textrm{(train)}$. 
Panels (a) and (b) give the distributions $P(\alpha)$ at $\varepsilon=0.5$ and $0.25$, respectively, for the indicated $U$ values. Panels (c) and (d) show 
$D_1$, $D_{\infty}$ 
versus $U$ at $\varepsilon=0.5$ and $0.25$, respectively. 
Panel (d) displays $d_\text{KL}(P^\textrm{NN},P^\textrm{ED})$ at the two scaled energies considered. The use of color bars, lines and symbols is the same as in  Figs.\ \ref{fig-1d-excited-states-alpha-Dq-KL-05-all} and  \ref{fig-1d-excited-states-alpha-Dq-KL-025-all}. 
} 
\label{fig-2d-excited-states-alpha-05-025-all}
\end{figure*}
We also illustrate $D_q$-training for excited states in the 2D BHH at non-integer filling, considering \revisionRAR{the} $4\times 4$ square lattice with $N=3$ particles, as we did for the ground state analysis in Sec.~\ref{sec:ground-state-2d}.
The NN results using linear and exponential activation functions with $q_+^\textrm{(train)}$ are shown in Fig.~\ref{fig-2d-excited-states-alpha-05-025-all} for scaled energies $\varepsilon=0.5$ and 
$\varepsilon=0.25$.
The predicted GFD in panels (c) and (d) reproduce fairly well the dependence of the exact $D_1$ and $D_\infty$ on $U$, with slightly larger fluctuations for strong interaction as is to be expected, but arguably even more accurately that in the 1D case. The evolution of the GFD reveals that the change in the excited eigenstate structure as a function of $U$ in 2D is more gentle than in the 1D chain. Accordingly, we see in panels (a) and (b) that the domain of the wave-function intensity distributions does not change significantly with $U$, keeping $\alpha\leqslant 3$ ($|\psi(f)|\geqslant 4\times10^{-5}$) even at $U=20.54$. 
The performance of the NN when predicting the $P(\alpha)$ distribution as a function of $U$ may seem only moderate, better for exponential activation, but note that the values of the KL divergence stay consistently below 0.5 up to $U\approx 10$, which is in agreement with the behaviour observed in 1D. Despite the absence of very small wave-function intensities, the distinct increase of KL for large interaction suggests that the NN training would \revisionRAR{similarly} benefit from the use of an activation function with an enhanced sensitivity for small Fock coefficients. 

\section{\label{sec:discussions-and conclusions} Discussions and Conclusions}

Let us briefly summarize what we have achieved thus far. We utilized a dense neural network (NN) (\HubNet\ \cite{Zhu2023}) to study interacting quantum many-body systems, with a focus on the 
Bose-Hubbard Hamiltonian (BHH). By just slightly improving on the learning rate and the optimizer---but without changing the structure of the NN---and training with respect to the energy, we show that \HubNet\ is capable of predicting the ground state energy $E_0$, as well as the wave-function $\ket{\Psi_0}$ both in 1D and 2D across four orders of magnitude 
of the interaction strength $U$. 
This means that one can get access, when given such a trained NN 
for the system sizes and densities considered, 
to the information of $E_0$ and $\ket{\Psi_0}$ at any interaction within the range $U\in[0.01,100]$ 
without the need to use any other numerical method, such as exact diagonalization. 
We emphasize that this includes $U$ values corresponding to states with completely different features. The NN furthermore enables us to capture important physical observables calculated from $\ket{\Psi_0}$. We showed as an example that the fractal dimensions $D_q$ and the wave-function intensity distribution 
can be well characterized across the whole $U$ range.

Excited-state training is notably more difficult. We find that training directly on average physical observables, 
i.e.,~the fractal dimensions $D_q$ 
\revisionRAR{\footnote{
Although the GFD are not quantum mechanical observables in the strict sense, they encode key structural information about the states that ultimately determine observable expectation values, especially in the vicinity of critical points. Hence, we anticipate that training on actual observables would lead to results comparable in predictive strength.}}
, offers a distinct advantage. This method, coupled with a physically motivated choice of the activation function, facilitates qualitatively accurate spectrum-wide predictions of the many-body state structure, as tested for scaled energies $\varepsilon=0.25$ and $0.5$, while bypassing the complexities of Gram-Schmidt towers. 
Ultimately, this provides a more scalable framework for estimating physical quantities without the overhead of explicit wave-function reconstruction.
In particular, we have successfully used this approach to describe the evolution of the excited state structure with $U$ across a chaotic phase. 

Looking toward the future, our results suggest a natural and distinct role for today's NN-based machine learning (ML) in quantum many-body physics. At present and for the as of today still relatively small system sizes accessible to NN approaches, ML methods are unlikely to be competitive---at least in the near term---with highly specialized and rigorously optimized techniques such as tensor-network algorithms \cite{Ostlund1995,Schollwck2011,Verstraete2023,Orus2014,Schuch2007} or modern implementations of exact diagonalization 
\cite{Troyer2005,Jung2020,Weinberg2019,Bollhfer2007,Balay1997,Balay2019,Hernandez2005,Roman2017,Pietracaprina2018,Nataf2014,Dabholkar2024}. 
These established methods are likely to remain indispensable for high-precision, large system size, benchmark-quality calculations.
However, as we have demonstrated here, current ML approaches offer a complementary strength: They can provide rapid and simultaneous access to 
energy and wave-function information of a medium-sized quantum many-body system across a broad range of physically relevant continuous parameters. Once trained, an NN effectively encodes this information in a compact form, enabling dense parameter sweeps over interaction strength, particle density, or other control parameters at negligible additional computational cost.

This capability opens a valuable use case. ML models can serve as an exploratory tool, allowing newcomers and experienced practitioners alike to obtain a quick, global overview of the bandwidth, phase structure, and qualitative physics exhibited by classes of quantum many-body systems, 
such as, e.g., Heisenberg, Hubbard, $tJ$, Haldane-Shastry and related models, without committing extensive resources upfront. 
Armed with this broad overview, as encoded in trained NNs, researchers can then identify physically interesting regimes and formulate more targeted questions. These focused problems can subsequently be addressed using bespoke, highly optimized numerical methods, such as DMRG/PEPS, that excel in restricted regions of parameter space. In this way, ML might not replace traditional approaches but instead act as a front-end guide, streamlining and sharpening the overall workflow of quantum many-body investigations.

\begin{acknowledgments}
We gratefully acknowledge funding by the University of Warwick's International Partnership Fund 2024 and support from Warwick's Scientific Computing Research Technology Platform (RTP) for its high-performance computing facilities and the Sulis Tier 2 HPC platform hosted by the RTP. Sulis is funded by EPSRC Grant EP/T022108/1 and the HPC Midlands+ consortium.
A.R.~acknowledges support through grant no. PID2024-156340NB-I00 funded by Ministerio de Ciencia, Innovaci\'on y Universidades/Agencia Estatal de Investigaci\'on (MICIU/AEI/10.13039/501100011033) and by the European Regional Development Fund (ERDF).
UK research data statement: ML codes for the adapted \HubNet\ are available at \url{https://github.com/DisQS/HubbardNet/tree/main}.
\end{acknowledgments}

%

\ifNOSUP\end{document}